\pdfoutput=1
\documentclass[iop]{emulateapj}

\slugcomment{}

\shorttitle{The physical environment around IRAS 17599-2148}
\shortauthors{L.~K. Dewangan et al.}

\begin{document}

\title{The physical environment around IRAS 17599-2148: infrared dark cloud and bipolar nebula}
\author{L.~K. Dewangan\altaffilmark{1}, D.~K. Ojha\altaffilmark{2}, I. Zinchenko\altaffilmark{3}, P. Janardhan\altaffilmark{1}, S.~K. Ghosh\altaffilmark{4}, and A. Luna\altaffilmark{5}}
\email{lokeshd@prl.res.in}
\altaffiltext{1}{Physical Research Laboratory, Navrangpura, Ahmedabad - 380 009, India.}
\altaffiltext{2}{Department of Astronomy and Astrophysics, Tata Institute of Fundamental Research, Homi Bhabha Road, Mumbai 400 005, India.}
\altaffiltext{3}{Institute of Applied Physics of the Russian Academy of Sciences, 46 Ulyanov st., Nizhny Novgorod 603950, Russia.}
\altaffiltext{4}{National Centre for Radio Astrophysics, Ganeshkhind, Pune - 411 007, India}
\altaffiltext{5}{Instituto Nacional de Astrof\'{\i}sica, \'{O}ptica y Electr\'{o}nica, Luis Enrique Erro \# 1, Tonantzintla, Puebla, M\'{e}xico C.P. 72840.}
\begin{abstract}
We present a multi-scale and multi-wavelength study to investigate the star formation process around IRAS 17599$-$2148 
that is part of an elongated filamentary structure (EFS) (extension $\sim$21 pc) seen in the {\it Herschel} maps. 
Using the {\it Herschel} data analysis, at least six massive clumps (M$_{clump}$ $\sim$777 -- 7024 M$_{\odot}$) 
are found in the EFS with a range of 
temperature and column density of $\sim$16--39~K and $\sim$0.6--11~$\times$~10$^{22}$ cm$^{-2}$ (A$_{V}$ $\sim$7--117 mag), respectively. 
The EFS hosts cold gas regions (i.e. infrared dark cloud) without any radio detection and a bipolar nebula (BN) linked with the H\,{\sc ii} region IRAS 17599$-$2148, tracing two distinct environments inferred 
through the temperature distribution and ionized emission. 
Based on virial analysis and higher values of self-gravitating pressure, the clumps are found unstable against gravitational collapse. 
We find 474 young stellar objects (YSOs) in the selected region and $\sim$72\% of these YSOs are found in the clusters 
distributed mainly toward the clumps in the EFS. 
These YSOs might have spontaneously formed due to processes not related to the expanding H\,{\sc ii} region.
At the edges of BN, four additional clumps are also associated with YSOs clusters, which appear to be influenced  by the expanding H\,{\sc ii} region. 
The most massive clump in the EFS contains two compact radio sources traced in the GMRT 1.28 GHz map 
and a massive protostar candidate, IRS~1 prior to an ultracompact H\,{\sc ii} phase. 
Using the VLT/NACO near-infrared images, IRS~1 is resolved with a jet-like feature within a 4200~AU scale.
 \end{abstract}
\section{Introduction}
\label{sec:intro}
% 
%The physics of star formation is complex. 
In recent years, {\it Spitzer} has revealed many massive star-forming (MSF) regions which contain 
the infrared dark clouds (IRDCs), mid-infrared (MIR) shells or bubbles, 
6.7 GHz methanol maser emission (MME), H\,{\sc ii} regions, and young star clusters together. 
It is indeed obvious that such sites host numerous complex physical processes of star formation. 
These sites offer to explore the key processes that govern the birth and feedback 
of massive stars ($\gtrsim$ 8 M$_{\odot}$) 
\citep[e.g.,][]{zin07,deharveng10,rathborne11,tackenberg12,tan14,beuther15,ragan15,dewangan15a,dewangan15b,xu16}. 
Furthermore, such regions also allow to understand the formation and evolution of stellar clusters.
In particular, the understanding of star formation processes demands a multi-scale and multi-wavelength approach.  
However, the study of close and large environments of such MSF regions is still lacking, 
which will help us to gain better insight into the ongoing physical processes. 

IRAS 17599$-$2148, situated at a distance of 4.2 kpc \citep{kim01}, 
has a bipolar appearance at wavelengths longer than 2 $\mu$m \citep{dewangan12}. 
The bipolar nebula was referred as two MIR bubbles \citep[CN107 and CN109;][]{churchwell07}. 
An H\,{\sc ii} region associated with IRAS 17599$-$2148 (i.e. G8.14+0.23 H\,{\sc ii} region) is 
excited by a single star of radio spectral class O6 \citep{kim01}. 
The 20 cm continuum peak \citep{dewangan12} and a Class~II 6.7 GHz 
MME \citep[peak velocity $\sim$19.83 km s$^{-1}$;][]{szymczak12} were 
reported toward the waist of the bipolar nebula \citep[see Figure~2 in][]{dewangan12}.  
Using the multi-wavelength analysis of the G8.14+0.23 H\,{\sc ii} region, 
\citet{dewangan12} found noticeable star formation activities on the periphery of the bipolar nebula and 
suggested an interaction of the H\,{\sc ii} region with its surroundings.
\citet{simon06} reported the IRDC candidates around IRAS 17599$-$2148 
(e.g. G008.25+00.16 and G008.20+00.18), which were identified based on the absorption 
features against the Galactic background in the 8.3 $\mu$m image. 
The integrated NANTEN $^{12}$CO intensity map (beam size $\sim2\farcm6$) of the cloud associated with IRAS 17599$-$2148 
was previously reported by \citet{takeuchi10}, 
which was referred as M008.2+0.2 (V$_{lsr}$ $\sim$19 km s$^{-1}$; a velocity range $\sim$12--25 km s$^{-1}$) 
in their work \citep[see Figure~14b in][]{takeuchi10}. Based on the $^{12}$CO gas distribution, the IRDC candidates traced in the 8.3 $\mu$m image 
are found to be physically linked with the molecular cloud M008.2+0.2 \citep[see Figure~15 in][]{takeuchi10}. However, an investigation of star formation activity in these IRDCs is not yet carried out. 
\citet{shirley13} reported the spectroscopic HCO$^{+}$ (3--2) 
line observations toward the 1.1 mm dust continuum sources in the IRAS 17599$-$2148 region 
and some of them are found toward the IRDCs and the bipolar nebula. 
These dust continuum sources are traced in a velocity range of 18 to 21 km s$^{-1}$, 
which are in agreement with the velocity of ionized gas \citep[i.e. 20.3 km s$^{-1}$;][]{kim01}. 
The knowledge of velocities of ionized gas, MME, and molecular gas confirms the 
physical association of the IRDC, MME, and bipolar nebula with 
the H\,{\sc ii} region linked with IRAS 17599$-$2148. 
Previous studies convincingly suggested that all these sources (i.e. H\,{\sc ii} region, IRDC, MME, 
and bipolar nebula) appear to be located in the same complex. 
Despite the availability of numerous observational data-sets, the understanding of close (below 10000 AU) 
and large (more than 10 pc) environments around the IRAS 17599$-$2148 region is missing. 
The ongoing physical processes around IRAS 17599$-$2148 are not systematically examined. 
In order to understand the physical environment and star formation processes around IRAS 17599$-$2148, we use the 
multi-wavelength data covering from the near-infrared (NIR) to radio wavelengths, 
which include the ESO Very Large Telescope (VLT) archival adaptive-optics NIR images and 
the Giant Metre-wave Radio Telescope (GMRT) radio continuum high-resolution maps at 0.61 and 1.28 GHz.  
The ESO-VLT NIR images are employed to probe the inner circumstellar environment of the 
infrared counterpart(s) (IRc) of exciting source(s) of the H\,{\sc ii} region and the 6.7 GHz MME. 
Our analysis is also focused to infer the distribution of dust temperature, column density, extinction, ionized emission, 
kinematics of molecular gas, and young stellar objects (YSOs). 
All together, such analysis helps us to characterize in depth the small and large-scale environments around IRAS 17599$-$2148, 
which will allow to understand the birth and feedback processes of massive stars as well as 
the formation and evolution of stellar clusters.

In Section~\ref{sec:obser}, we provide the details of the multi-wavelength datasets used in this paper. 
In Section~\ref{sec:data}, we characterize the physical environment of the region around IRAS 17599$-$2148. 
Furthermore, we identify the stellar populations around the IRAS 17599$-$2148 region and infer their clustering. 
In Section~\ref{sec:disc}, we discuss the possible star formation scenario based on our findings. 
Finally, in Section~\ref{sec:conc}, we provide a summary of conclusions.
\section{Data and analysis}
\label{sec:obser}
Mutli-scale and multi-wavelength data have been used to explore the physical environments around 
IRAS 17599$-$2148. In this paper, we selected a region of $\sim$20$\farcm8$ $\times$ 13$\farcm2$ 
($\sim$25.4 pc $\times$ 16.1 pc at a distance of 4.2 kpc), centered at $\alpha_{2000}$ = 18$^{h}$02$^{m}$54.5$^{s}$, 
$\delta_{2000}$ = $-$21$\degr$46$\arcmin$33$\farcs6$.
\subsection{New Observations}
\subsubsection{Radio Continuum Observations}
Radio continuum observations of IRAS 17599$-$2148 at 0.61 and 1.28 GHz bands were obtained using 
the GMRT on 2012 Dec 29 and 2012 Dec 23 (Proposal Code: 23\_054; PI: L.~K. Dewangan), respectively. 
The radio data reduction was performed using the AIPS software, following similar method as described in \citet{mallick12,mallick13}. 
The 0.61 and 1.28 GHz maps have rms noise of 0.34 and 0.38 mJy/beam, respectively.
The synthesized beam sizes of the final 0.61 and 1.28 GHz maps are $\sim$5\farcs6 $\times$ 5\farcs2 
and $\sim$ 2\farcs8 $\times$ 2\farcs4, respectively.

Note that a large amount of emission from the Galactic plane was expected during the observations, 
which influences the low frequency observations and increases the effective antennae temperature. 
Since, IRAS 17599$-$2148 is located in the Galactic plane, 
our GMRT 0.61 and 1.28 GHz maps are corrected for system temperature \citep[see][]{omar02,mallick12,mallick13,baug15}. 
The details about the adopted procedure of system temperature corrections are described in \citet{mallick12,mallick13}. 
\subsection{Archival data}
\subsection{NIR images}
\subsubsection{Adaptive-optics NIR images}
The imaging observations of IRAS 17599$-$2148 were taken with 8.2m VLT with 
NAOS-CONICA (NACO) adaptive-optics system \citep{lenzen03,rousset03} 
in H-band ($\lambda _{c}=1.66\, \mu \rm m, \Delta \lambda =0.33\, \mu \rm m$), 
K$_{s}$-band ($\lambda _{c}=2.18\, \mu \rm m, \Delta \lambda =0.35\, \mu \rm m$) and 
L$^{\prime}$-band ($\lambda _{c}=3.80\, \mu \rm m, \Delta \lambda =0.62\, \mu \rm m$). 
We downloaded these data from the ESO-Science Archive Facility 
(ESO proposal ID: 089.C-0455(A); PI: Jo\~{a}o Alves). 
The single frame exposure time was 21 s in L$^{\prime}$-band, and 24 s in H- and K$_{s}$-bands. 
We used five H frames, five K$_{s}$ frames, and eleven L$^{\prime}$ frames. 
In each band, the final NACO image was obtained through the standard analysis procedure such as sky subtraction, 
image registration, combining with median method, and astrometric calibration, using IRAF and STAR-LINK softwares. 
The astrometric calibration of the NACO images was done using the UKIRT NIR Galactic Plane Survey \citep[GPS;][]{lawrence07} 
K-band point sources. 
The VLT/NACO H, K$_{s}$, and L$^{\prime}$ images have resolutions of 0\farcs2 ($\sim$840 AU), 0\farcs2 ($\sim$840 AU), and 
0\farcs1 ($\sim$420 AU), respectively.
\subsubsection{UKIDSS GPS NIR Data}
Deep NIR photometric {\it JHK} magnitudes of point sources have been retrieved from the UKIDSS GPS sixth archival data release (UKIDSSDR6plus). 
The UKIDSS observations (resolution $\sim$$0\farcs8$) were carried out using the UKIRT Wide Field Camera 
\citep[WFCAM;][]{casali07}. The UKIDSS GPS photometric data were calibrated using the Two Micron All Sky Survey \citep[2MASS;][]{skrutskie06} data. 
We extracted only reliable NIR photometric data following the recommendations given in \citet{lucas08} and \citet{dewangan15b}. 
To avoid saturation, the sources were obtained fainter than J = 12.1, H = 11.1, and K = 10.0 mag in our selected GPS catalog. 
Furthermore, the 2MASS photometric magnitudes were retrieved for bright sources that were saturated in the GPS catalog. 
\subsection{{\it Spitzer} Data}
Photometric images and magnitudes of point sources have been obtained from 
the {\it Spitzer}-Galactic Legacy Infrared Mid-Plane Survey Extraordinaire 
\citep[GLIMPSE;][]{benjamin03} survey at 3.6--8.0 $\mu$m (resolution $\sim$2$\arcsec$). 
The photometric magnitudes were extracted from the GLIMPSE-I Spring '07 highly reliable catalog. 
Additionally, we also obtained photometry of point sources from \citet{dewangan12} near the 
bipolar nebula, which are not listed in the GLIMPSE catalog \citep[see][]{dewangan12}. 
The MIPS Inner Galactic Plane Survey \citep[MIPSGAL;][]{carey05} 24 $\mu$m data have been utilized in this paper. 
The MIPSGAL 24 $\mu$m photometry has also been obtained from \citet{gutermuth15}. 
We also performed aperture photometry in the MIPSGAL 24 $\mu$m image to extract point sources, which 
are not listed in the MIPSGAL 24 $\mu$m catalog.
The photometry was extracted using a $7\arcsec$ aperture radius and a sky annulus from $7\arcsec$ to $13\arcsec$ 
in IRAF \citep[e.g.][]{dewangan15b}. 
The MIPS zero-magnitude flux density, including aperture correction, was adopted for the photometric calibration, as listed in the MIPS 
Instrument Handbook-Ver-3. 
\subsection{{\it Herschel}, ATLASGAL, and SCUBA-2 Data}
Far-infrared (FIR) and sub-millimeter (sub-mm) data adopted in this work have been retrieved from the 
{\it Herschel} Space Observatory data archives. The processed level2$_{-}$5 images 
at 70 $\mu$m, 160 $\mu$m, 250 $\mu$m, 350 $\mu$m, and 500 $\mu$m were 
obtained using the {\it Herschel} Interactive Processing Environment \citep[HIPE,][]{ott10}. 
The beam sizes of these images are 5$\farcs$8, 12$\arcsec$, 18$\arcsec$, 25$\arcsec$, 
and 37$\arcsec$ \citep{poglitsch10,griffin10}, respectively. 

The Submillimetre Common-User Bolometer Array 2 (SCUBA-2) 
continuum map at 850 $\mu$m (beam size $\sim$14$\arcsec$) was obtained from the James 
Clerk Maxwell Telescope (JCMT) science archive. In our selected region,  we also obtained the positions of fourteen dust clumps at 870 $\mu$m 
\citep{contreras13a} from the APEX Telescope Large Area Survey of the Galaxy \citep[ATLASGAL; beam size $\sim$19$\farcs$2;][]{schuller09}.
\subsection{CO (J=3$-$2) Line Data}
The JCMT HARP CO (3-2) processed data cube was downloaded and the spatial resolution of the data cube is 16$\arcsec$. 
The CO observations were observed on 11 June 2009. 
The data have a typical rms sensitivity (1$\sigma$) of $\approx0.3$~K.
\section{Results}
\label{sec:data}
\subsection{Large-scale environment around IRAS 17599$-$2148}
\label{subsec:u1}
The distribution of molecular gas, dense materials, and ionized emission enables us to infer 
physical environment present in and around the IRAS 17599$-$2148 region. 
\subsubsection{Infrared and molecular emission}
\label{subsec:u16}
\begin{figure*}
\epsscale{0.93}
\plotone{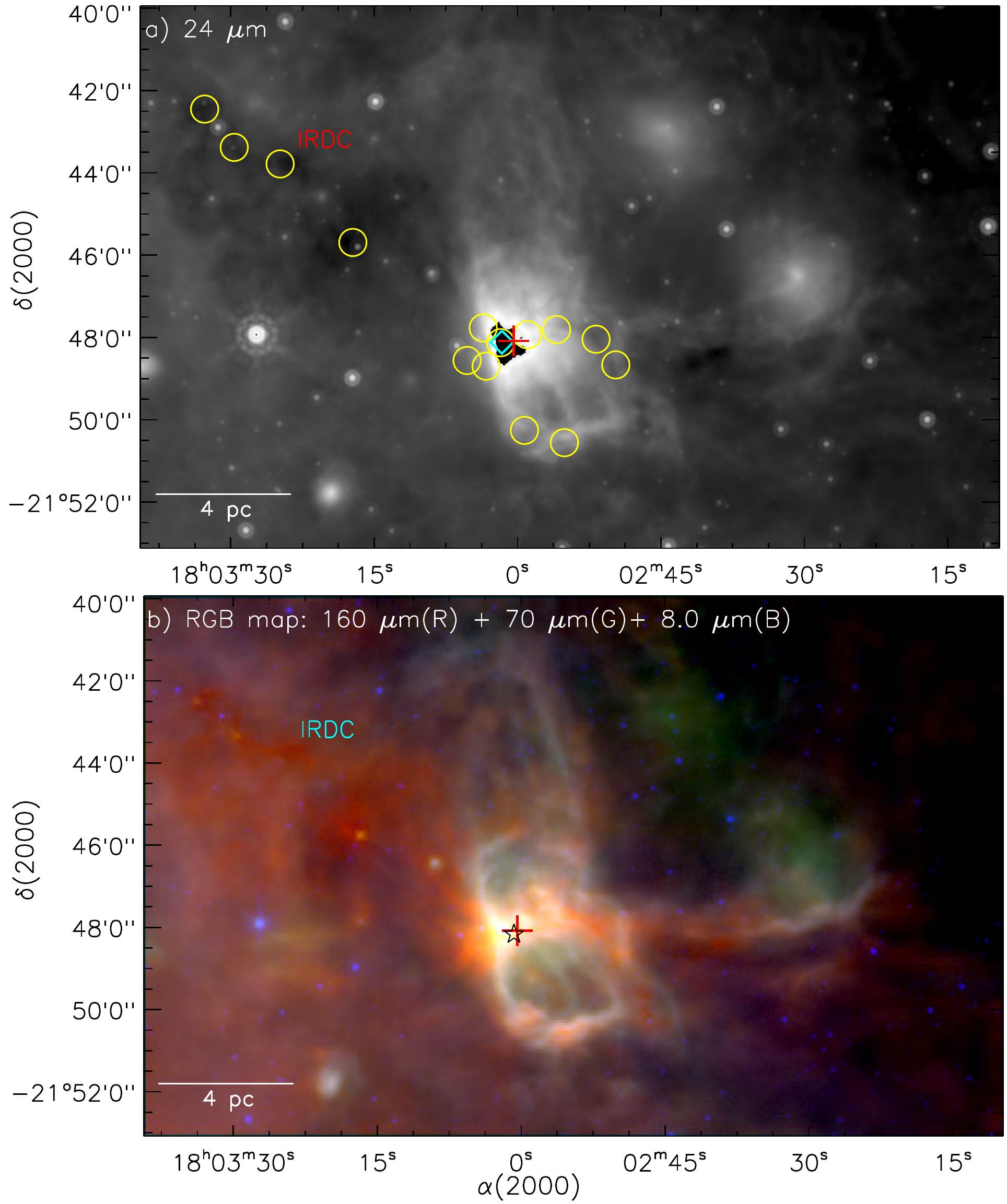}
\caption{\scriptsize Distribution of the MIR and FIR emission toward the region around IRAS 17599$-$2148 
(size of the selected region $\sim$20$\farcm8$ $\times$ 13$\farcm2$ 
($\sim$25.4 pc $\times$ 16.1 pc at a distance of 4.2 kpc); centered at $\alpha_{2000}$ = 18$^{h}$02$^{m}$54.5$^{s}$, 
$\delta_{2000}$ = $-$21$\degr$46$\arcmin$33$\farcs6$).
a) The MIPSGAL 24 $\mu$m image of the region probed in this paper. A diamond symbol shows the  VLA 6 cm (5 GHz) radio detection \citep{becker94}. The 24 $\mu$m image is saturated near the radio detection. The ATLASGAL clumps at 870$\mu$m \citep{contreras13a} are also marked by yellow circles. 
b) The image is the result of the combination of three bands: 160 $\mu$m (red), 70 $\mu$m (green), 
and 8.0 $\mu$m (blue). A star symbol indicates the position of the 6.7-GHz MME. 
In both the panels, the position of IRAS 17599$-$2148 (+) is marked. 
The scale bar corresponding to 4 pc (at a distance of 4.2 kpc) is shown in both panels. 
The images have revealed an IRDC \citep{simon06} and a bipolar nebula \citep{dewangan12}.}
\label{fig1}
\end{figure*}

\begin{figure*}
\epsscale{0.75}
\plotone{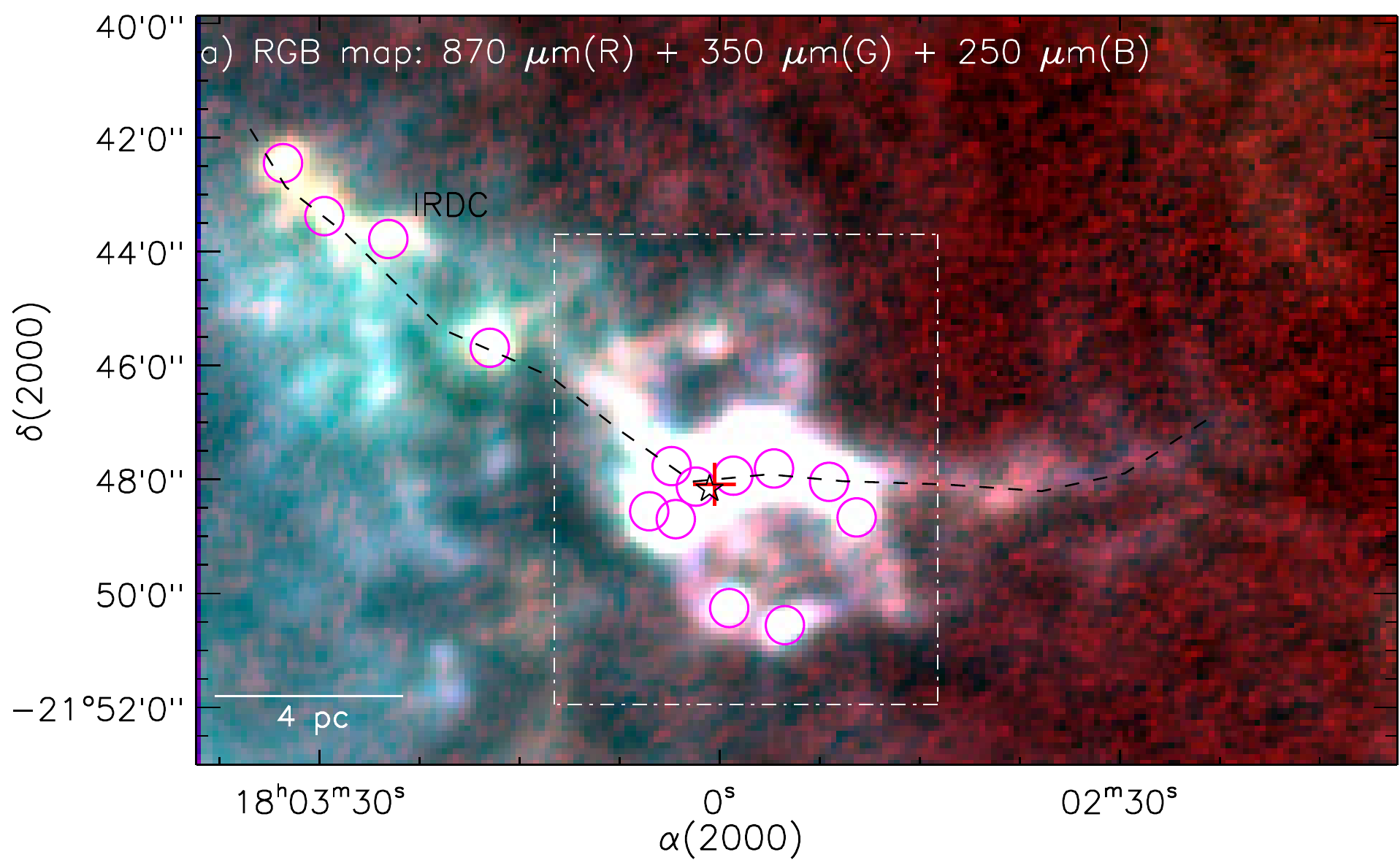}
\epsscale{0.49}
\plotone{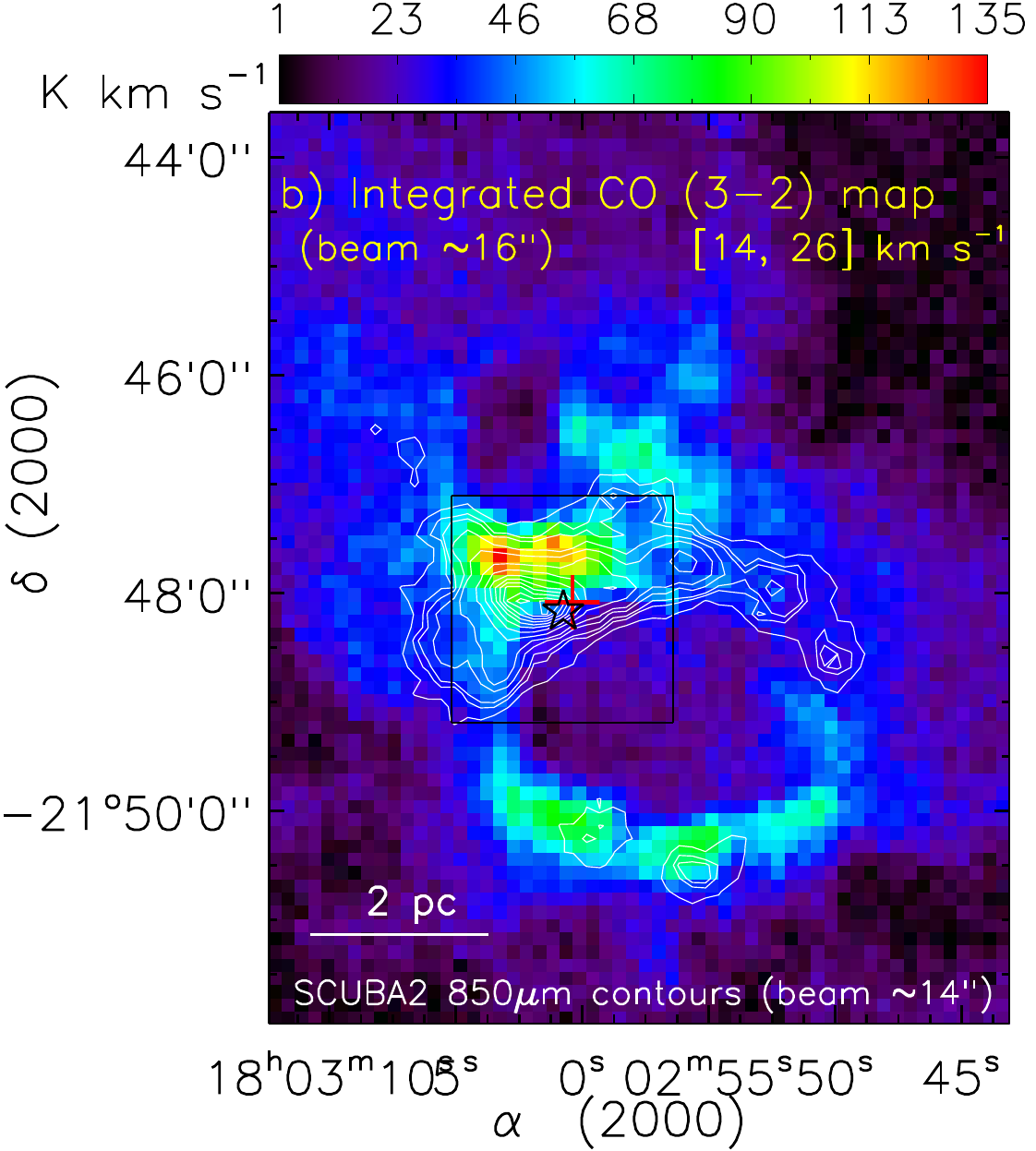}
\caption{\scriptsize a) Three-color composite image ({\it red}, ATLASGAL 870 $\mu$m; {\it green}, {\it Herschel} 350 $\mu$m; 
{\it blue},{\it Herschel} 250 $\mu$m; in linear scale) 
of the region around IRAS 17599$-$2148, which clearly traces the cold dust emission. An elongated filamentary structure is highlighted by a dashed curve. 
The ATLASGAL clumps at 870$\mu$m \citep{contreras13a} are also marked by magenta circles. 
The faint dust emissions (or infrared bridges) traced in the {\it Herschel} images are also found between the ATLASGAL clumps, 
indicating that this chain of dense clumps/cores belongs to a single entity (i.e., elongated filamentary structure; see a dashed curve) (also see text for details). The dotted-dashed white box encompasses the area shown in Figure~\ref{fig2}b. 
The scale bar at the bottom-left corner corresponds to 4 pc (at a distance of 4.2 kpc). 
b) Integrated CO (3--2) emission map overlaid with the SCUBA-2 850 $\mu$m contours. 
The 850 $\mu$m contours are superimposed with levels of 4, 8, 10, 15, 20, 30, 40, 50, 60, 70, 80, 90, and 97\% of 
the peak value (i.e., 15.5 mJy/arcsec$^{2}$). 
The figure shows the distribution of molecular gas and dense materials in the region around IRAS 17599$-$2148. 
The solid black box encompasses the area shown in Figure~\ref{fig3}. 
The scale bar at the bottom-left corner corresponds to 2 pc (at a distance of 4.2 kpc).
In both the panels, the marked symbols are similar to those shown in Figure~\ref{fig1}.}
\label{fig2}
\end{figure*}

On a larger scale, an IRDC and a bipolar nebula, associated with the bright extended emission, 
are visually observed in the MIPSGAL 24 $\mu$m image (see Figure~\ref{fig1}a).
In the MIPSGAL 24 $\mu$m image, the embedded stellar sources are also seen toward the IRDC. 
Fourteen ATLASGAL dust continuum sources at 870 $\mu$m are also marked in Figure~\ref{fig1}a. 
The ATLASGAL sources are found toward the IRDC and the bipolar nebula, tracing cold and dense regions. 
It appears as a chain of dense clumps/cores in our selected target region \citep[e.g.,][]{tafalla15}.
In Figure~\ref{fig1}a, the 24 micron image is saturated near the IRAS position, where the radio 
continuum peak position was reported \citep{becker94,dewangan12}. 
Figure~\ref{fig1}b shows a three-color composite image made using {\it Herschel} 160 $\mu$m in red, {\it Herschel} 70 $\mu$m in green, 
and GLIMPSE 8.0 $\mu$m in blue. The IRDC appears as bright emission regions at wavelengths longer than 24 $\mu$m. 
The bipolar nebula is also seen as the brightest extended emission in Figure~\ref{fig1}b. 
Figure~\ref{fig2}a shows the ATLASGAL and {\it Herschel} sub-mm images of the region, which trace the cold dust emission 
(see Section~\ref{subsec:temp} for quantitative estimate). The bipolar nebula is traced in all the {\it Herschel} images. 
Interestingly, we infer an elongated filamentary morphology (extension $\sim$21 pc) in the ATLASGAL and {\it Herschel} sub-mm images, 
where several condensations are found. Furthermore, the IRDC and the G8.14+0.23 H\,{\sc ii} region (including the bipolar nebula) 
are part of the filamentary structure (see a dashed curve in Figure~\ref{fig2}a), which is embedded in the cloud associated 
with IRAS 17599$-$2148 \citep[i.e. M008.2+0.2; see Figures~14b and~15 in][]{takeuchi10}. 
The integrated NANTEN $^{12}$CO intensity map (beam size $\sim2\farcm6$) traces a continuous velocity structure in the direction of IRAS 17599$-$2148 \citep{takeuchi10}. 
It includes the eastern and western sides of the bubble as well as the bubble itself 
\citep[i.e., entire elongated filamentary structure; see Figure~14b in][]{takeuchi10}.
Hence, this previous molecular line data, even with coarse beam size, further confirms the existence of the elongated filamentary structure. 
In Figure~\ref{fig2}a, the {\it Herschel} images trace the faint dust emission (or infrared bridges) between the ATLASGAL dust continuum 
sources, which are located within the elongated filamentary structure. 
Taken together, the sub-mm images and the previously published CO line data reveal the existence of a single elongated filamentary 
structure in the region probed in this paper.

In Figure~\ref{fig2}b, we present the JCMT CO(J = 3--2) gas emission in the direction of the IRAS position. 
Note that we do not have JCMT molecular line observations for the entire field of view, which is probed in this paper. 
The JCMT CO data are available only toward the bipolar nebula. 
Considering the better resolution of the JCMT HARP CO(3-2) data compared to the previously published 
NANTEN $^{12}$CO data, we are able to observe the bipolar morphology in the integrated JCMT CO intensity map (in a velocity range of 14--26 km s$^{-1}$). 
The kinematics of molecular gas are described in Section~\ref{sec:coem}.
In Figure~\ref{fig2}b, the 850 $\mu$m continuum emission, which traces the densest materials, 
is also overlaid on the integrated CO map and are mostly distributed toward the waist of the bipolar nebula. 
The 850 $\mu$m continuum map (beam size $\sim$14$\arcsec$) reveals two peaks (i.e. dense clumps/cores) 
and one of them coincides with a Class~II 6.7-GHz MME. The 6.7 GHz MME is a reliable tracer of massive young stellar 
objects (MYSOs) \citep[e.g.][]{walsh98,minier01,urquhart13}. 
\subsubsection{Radio continuum emission}
\label{subsec:radio}
\begin{figure*}
\epsscale{0.65}
\plotone{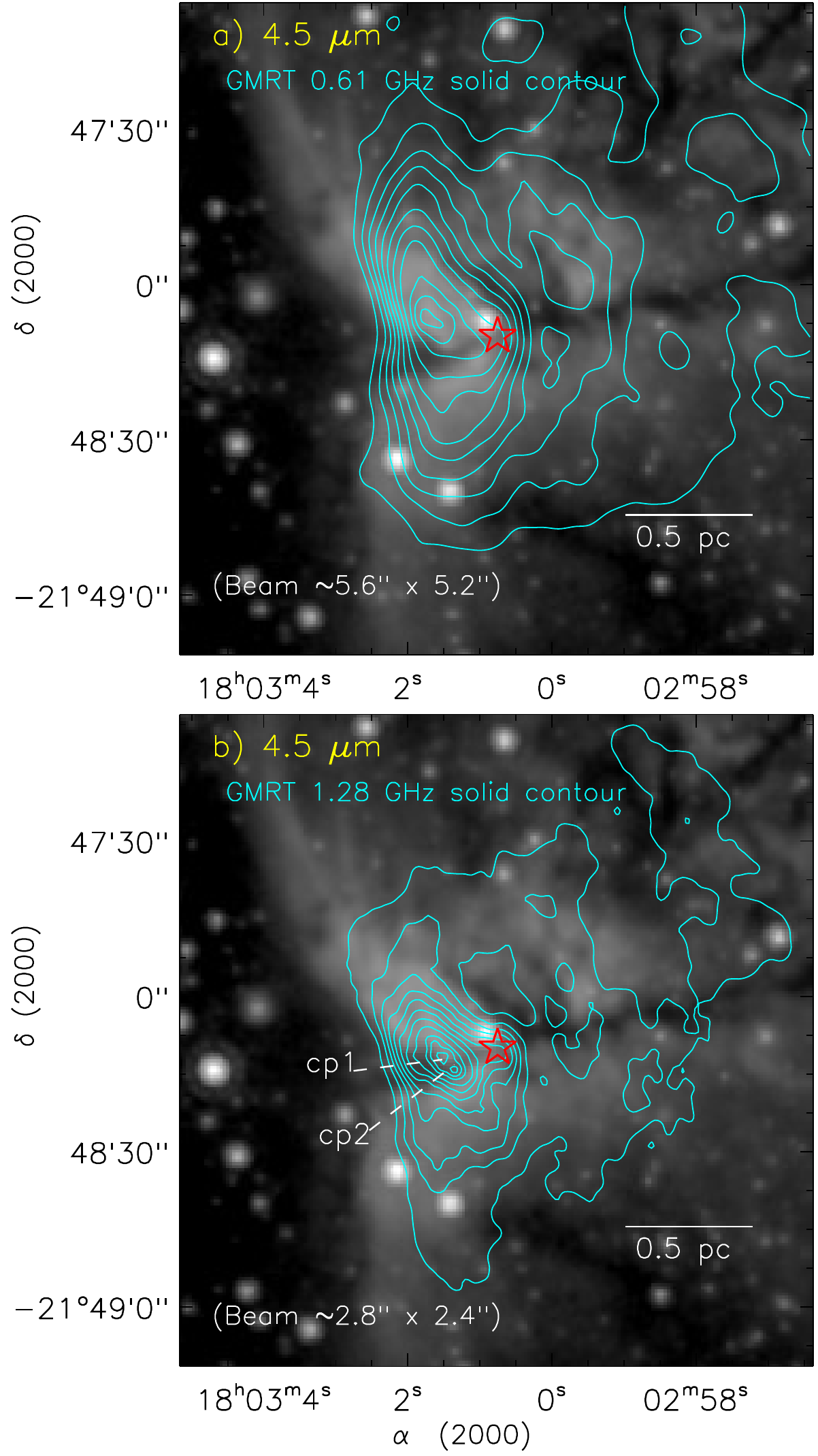}
\caption{\scriptsize High-resolution GMRT radio continuum emissions at 0.61 GHz (beam size $\sim$5\farcs6 $\times$ 5\farcs2) (a)
and 1.28 GHz (beam size $\sim$2\farcs8 $\times$ 2\farcs4) (b). In each panel, the radio contours are overlaid on the 4.5 $\mu$m image. 
The 1.28 GHz map traces two compact peaks (i.e. cp1 and cp2), which are also labeled in the bottom panel. 
In both the panels, the contours are overlaid with levels of 10, 20, 30, 40, 50, 60, 70, 80, 90, 95, and 99\% of the peak values 
(i.e. 0.0738 Jy/beam for 0.61 GHz; 0.0225 Jy/beam for 1.28 GHz). In both the panels, a star symbol indicates the position of the 6.7-GHz MME. 
In both the panels, the scale bar at the bottom-right corner corresponds to 0.5 pc (at a distance of 4.2 kpc).}
\label{fig3}
\end{figure*}

In order to trace the distribution of ionized emission, we present 
high-resolution GMRT radio continuum maps at 0.61 GHz (beam size $\sim$5\farcs6 $\times$ 5\farcs2) 
and 1.28 GHz (beam size $\sim$2\farcs8 $\times$ 2\farcs4) of the region around IRAS 17599$-$2148. 
Figure~\ref{fig3} shows the GMRT 0.61 (top panel) and 1.28 GHz (bottom panel) contours overlaid on the {\it Spitzer} 4.5 $\mu$m image.  
Note that the radio emission is not detected toward the IRDC (not shown here) and is found only toward the previously 
known G8.14+0.23 H\,{\sc ii} region. 
However, these new maps provide more insight into the radio peak, due to high-resolution map, 
compared to previously reported radio continuum map at 20 cm (beam $\sim$6$\arcsec$) \citep{dewangan12}. 
Both the GMRT maps show the similar morphology of the ionized emission and their peak positions are offset from the 6.7-GHz MME. 
The extended morphology and a single peak at 0.61 GHz map are similar as previously 
seen at 20 cm map \citep{dewangan12}. 
In addition to the extended morphology, due to high-resolution observations, the 1.28 GHz map has 
resolved a single peak (as seen in the 0.61 GHz map) into two compact peaks (see Figure~\ref{fig3}b) 
(i.e., cp1 ($\alpha_{2000}$ = 18$^{h}$03$^{m}$01$^{s}$.49, $\delta_{2000}$ = $-$21$\degr$48$\arcmin$12$\arcsec$.23; 
peak flux density = 22.52 mJy/beam; angular size =28.78$\arcsec$ $\times$ 13.95$\arcsec$; 
deconvolved size = 28.66$\arcsec$ $\times$ 13.71$\arcsec$); 
cp2 ($\alpha_{2000}$ = 18$^{h}$03$^{m}$01$^{s}$.38, $\delta_{2000}$ = $-$21$\degr$48$\arcmin$13$\arcsec$.82; 
peak flux density = 22.63 mJy/beam; angular size =14.25$\arcsec$ $\times$ 9.93$\arcsec$; 
deconvolved size = 14.02$\arcsec$ $\times$ 9.57$\arcsec$)). 
Using the 1.28 GHz map, we computed the integrated flux densities equal to 1355.50 mJy and 480.12 mJy 
for cp1 and cp2 peaks, respectively. 
These integrated flux densities are utilized to compute the number of Lyman continuum photons (N$_{uv}$).
The equation of N$_{uv}$ is given by \citep{matsakis76}:
\begin{equation}
%\begin{split}
N_{uv} (s^{-1}) = 7.5\, \times\, 10^{46}\, \left(\frac{S_{\nu}}{Jy}\right)\left(\frac{D}{kpc}\right)^{2} 
\left(\frac{T_{e}}{10^{4}K}\right)^{-0.45} \\ \times\,\left(\frac{\nu}{GHz}\right)^{0.1}
%\end{split}
\end{equation}
\noindent where S$_{\nu}$ is the measured total flux density in Jy, D is the distance in kpc, 
T$_{e}$ is the electron temperature, and $\nu$ is the frequency in GHz. 
The analysis is done for a distance of 4.2 kpc and for the electron temperature of 10000~K.
We find N$_{uv}$ (or logN$_{uv}$) to be $\sim$1.86 $\times$ 10$^{48}$ s$^{-1}$ (48.27) and $\sim$6.6 $\times$ 10$^{47}$ s$^{-1}$ (47.82) 
for cp1 and cp2, respectively. 
These estimates correspond to a single ionizing star of spectral 
type O8V--O8.5V and O9V--O9.5V (see Table 1 in \citet{martins05} for theoretical values) for cp1 and cp2, respectively.  
Based on these results, within a 18$\arcsec$ of the IRAS position, two radio O-spectral type sources and a 6.7 GHz MME (without radio peak) 
are present, indicating the presence of different early evolutionary stages of massive star formation. 
The search of the driving sources of the radio peaks and the 6.7 GHz MME is presented in Section~\ref{sec:small}.

We also computed the integrated flux densities of the extended radio morphology seen (at 3-$\sigma$) in both the radio maps. 
Using equation~1, the estimation of N$_{uv}$ is also performed for both the radio frequencies (0.61 GHz and 1.28 GHz) separately. 
Taking into account, D = 4.2 kpc, T$_{e}$ = 10000 K, and S$_{1.28}$ = 3.39 Jy, 
we find N$_{uv}$ (or logN$_{uv}$) = 4.7 $\times$ 10$^{48}$ s$^{-1}$ (48.67).
Similarly, we obtain N$_{uv}$ (or logN$_{uv}$) = 6.9 $\times$ 10$^{48}$ s$^{-1}$ (48.84) for D, T$_{e}$, and S$_{0.61}$ = 5.39 Jy. 
The estimates of N$_{uv}$ values at different frequencies correspond to a single ionizing star of O6V--O6.5V spectral type \citep{martins05}, 
which is consistent with the previously reported spectral type of the ionizing source of the G8.14+0.23 H\,{\sc ii} region \citep[e.g.][]{kim01}.  

Within the elongated filamentary structure, the radio emission is detected toward the IRAS 17599$-$2148 position and is absent toward the IRDC, 
revealing different evolutionary stages of star formation.
\subsection{{\it Herschel} temperature and column density maps}
\label{subsec:temp}
In recent years, the {\it Herschel} temperature and column density maps have been very useful tool for studying 
the distribution of column density, temperature, extinction, and clump mass in a given star-forming region.
In this section, we also present these maps of the selected region around IRAS 17599$-$2148. 
We produced the temperature and column density maps of the region, 
in a manner similar to that described in \citet{mallick15} \citep[also see][]{dewangan15b}. 
Here, we also give a brief step-by-step description of the procedures. 

The {\it Herschel} temperature and column density maps are obtained from a  pixel-by-pixel spectral energy distribution (SED) 
fit with a modified blackbody to the cold dust emission in the {\it Herschel} 160--500 $\mu$m wavelengths. 
The {\it Herschel} 70 $\mu$m data are not included in the analysis, because the 70 $\mu$m emission is dominated by UV-heated warm dust. 
The {\it Herschel} image at 160 $\mu$m is in the units of Jy pixel$^{-1}$ and 
the images at 250--500 $\mu$m are calibrated in the surface brightness unit of MJy sr$^{-1}$. 
The plate scales of the 160, 250, 350, and 500 $\mu$m images are 6.4, 6, 10, and 14 arcsec/pixel, respectively.  
Before the SED fit, all images were convolved to the angular resolution of 500 $\mu$m image ($\sim$37$\arcsec$) and were brought to the 
same flux unit (i.e. Jy pixel$^{-1}$). Furthermore, the images were regridded  to the pixel size of 
500 $\mu$m image ($\sim$14$\arcsec$). These procedures were performed using the convolution kernels available in the HIPE software. 
Next, the sky background flux level was determined to be 0.393, 1.125, 2.528, and $-$0.186 Jy pixel$^{-1}$ for the 500, 350, 250, and 
160 $\mu$m images (size of the selected region $\sim$7$\farcm$7 $\times$ 8$\farcm$2; 
centered at: $\alpha_{J2000}$ = 18$^{h}$03$^{m}$41$^{s}$, $\delta_{J2000}$ = $-$20$\degr$52$\arcmin$49.6$\arcsec$), respectively.
The featureless dark area away from the selected target region, to avoid diffuse emission associated with the target, 
was carefully selected for the background estimation. 

\begin{figure*}
\epsscale{0.59}
\plotone{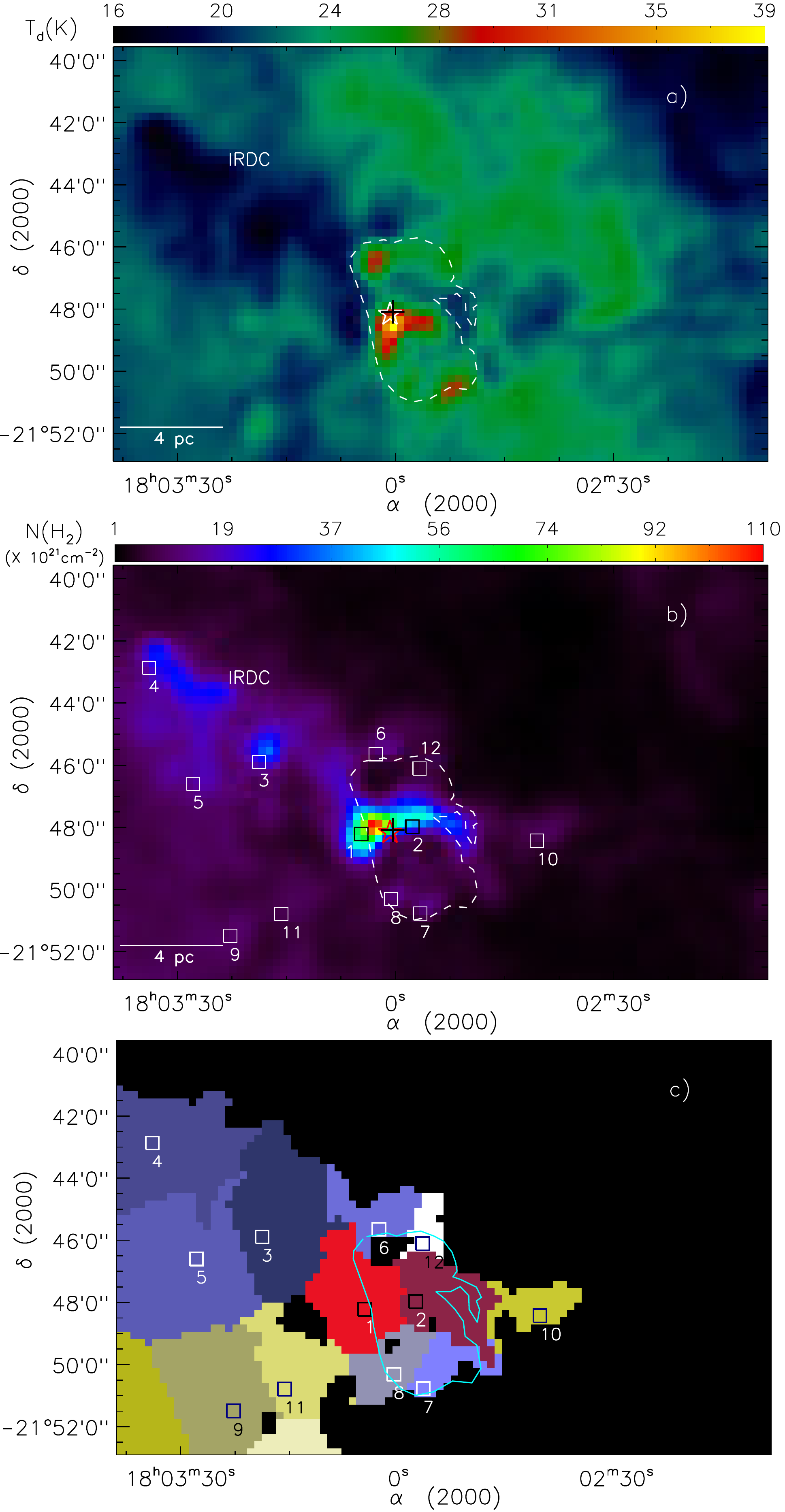}
\caption{\scriptsize {\it Herschel} temperature map (a) and column density ($N(\mathrm H_2)$) map (b) of the region around IRAS 17599$-$2148 
(see text for details). The column density map allows to infer the extinction with $A_V=1.07 \times 10^{-21}~N(\mathrm H_2)$ 
and to identify the clumps (see text for details). The identified clumps are marked by square symbols and the boundaries of these 
clumps are shown in Figure~\ref{fig4}c (also see Table~\ref{tab1}). 
c) The boundary of each identified clump is highlighted along with its corresponding clump ID (see Table~\ref{tab1} and also Figure~\ref{fig4}b). 
A bipolar nebula is depicted from the {\it Spitzer} 8 $\mu$m emission \citep[see][]{dewangan12}, which is shown in all the panels. 
In the first two panels, the other marked symbols are similar to those shown in Figure~\ref{fig1}. 
In the first two panels, the scale bar at the bottom-right corner corresponds to 4 pc (at a distance of 4.2 kpc). }
\label{fig4}
\end{figure*}

In the final step, a modified blackbody was fitted to the observed fluxes on a pixel-by-pixel basis, 
in order to obtain the maps \citep[see equations 8 and 9 given in][]{mallick15}. 
The fitting was performed using the four data points for each pixel, retaining the 
dust temperature (T$_{d}$) and the column density ($N(\mathrm H_2)$) 
as free parameters. 
In the calculations, we utilized a mean molecular weight per hydrogen molecule ($\mu_{H2}$=) 2.8 
\citep{kauffmann08} and an absorption coefficient ($\kappa_\nu$ =) 0.1~$(\nu/1000~{\rm GHz})^{\beta}$ cm$^{2}$ g$^{-1}$, 
including a gas-to-dust ratio ($R_t$ =) of 100, with a dust spectral index of $\beta$\,=\,2 \citep[see][]{hildebrand83}. 
The final temperature and column density maps (resolution $\sim$37$\arcsec$) are shown in Figure~\ref{fig4}.

The {\it Herschel} temperature map traces the IRDC in a temperature range of about 16--22~K, 
while the considerably warmer gas (T$_{d}$ $\sim$25-39 K) is found toward the IRAS position 
(including the H\,{\sc ii} region and the bipolar nebula; see Figure~\ref{fig4}a).
It implies that the cold gas is exclusively found toward the IRDC, while the warmer gas is detected toward the H\,{\sc ii} region. 
In Figure~\ref{fig4}b, the IRDC and the bipolar nebula are traced with higher column densities. 
Several condensations are seen in the column density map (see Figure~\ref{fig4}b). 
The condensation having the highest column density 
(peak $N(\mathrm H_2)$ $\sim$1.1~$\times$~10$^{23}$ cm$^{-2}$; A$_{V}$ $\sim$117 mag) is 
associated with the H\,{\sc ii} region and the 6.7 GHz MME. 
The relation between optical extinction and hydrogen column density 
\citep[$A_V=1.07 \times 10^{-21}~N(\mathrm H_2)$;][]{bohlin78} is adopted here. 
\subsubsection{{\it Herschel} clumps}
\label{subsec:clm}
In the column density map, the {\it clumpfind} algorithm \citep{williams94} has been applied to identify the clumps 
and their total column densities. 
Twelve clumps are identified, which are labeled in Figure~\ref{fig4}b and their boundaries are also shown in Figure~\ref{fig4}c. 
We estimated the masses of these clumps using their total column densities. 
The mass of a single clump can be computed using the formula:
\begin{equation}
M_{clump} = \mu_{H_2} m_H Area_{pix} \Sigma N(H_2)
\end{equation}
where $\mu_{H_2}$ is assumed to be 2.8, $Area_{pix}$ is the area subtended by one pixel, and 
$\Sigma N(\mathrm H_2)$ is the total column density. 
The mass of each {\it Herschel} clump is listed in Table~\ref{tab1}. 
The table also contains an effective radius of each clump, which is provided by the {\it clumpfind} algorithm. 
The clump masses vary between 410 M$_{\odot}$ and 7024 M$_{\odot}$. 
The elongated filamentary structure is associated with at least six massive clumps (i.e. IDs 1--5 and 10), 
indicating the fragmentation of the cloud. 
The most massive condensation (i.e ID 1) hosts two compact radio continuum sources and the 6.7 GHz MME, which is located at the waist of the bipolar nebula. 
At least three massive and cold condensations (i.e. IDs 3--5) are seen toward the IRDC. 
At least four condensations (i.e. IDs 6--8 and 12) are also found at the edges of the bipolar nebula. 

In order to examine the internal dynamical properties of these clumps in more detail, we computed their physical parameters 
(such as, spherical free-fall time, virial mass, virial parameter, self-gravitating pressure, and Mach number). 
Using mass ($M_{clump}$) and radius (R$_{c}$) of each clump,  we can estimate the mean mass density $\rho_{0}$, and 
the spherical free-fall time, ($t_{ff,sph}$) = $(3\pi/ 32G\rho_{0}$)$^{1/2}$ (= $16.6 (R_{c}/pc)^{3/2} (M_{clump}/M_\odot)^{-1/2}$ Myr). 
We estimated $t_{ff,sph}$ values for our identified clumps range from 0.4 to 0.77 Myr, which are tabulated in Table~\ref{tab1}. 

In each clump, the pressure exerted by the self-gravity of the molecular gas can be estimated using 
$P_{clump}$ $\approx$ $\pi G (M_{clump}/ \pi R_{c}^2$)$^2$. The value of $P_{clump}$ is tabulated in Table~\ref{tab1}. 
In general, the pressure associated with a typical cool molecular cloud ($P_{MC}$) is found to be 
$\sim$10$^{-11}$--10$^{-12}$ dynes cm$^{-2}$ (for a temperature $\sim$20 K 
and the particle density $\sim$10$^{3}$--10$^{4}$ cm$^{-3}$) \citep[see Table 7.3 of][]{dyson80}, which is much 
lower than $P_{clump}$. This comparison indicates an acting of self-gravity and a signature of star formation activity in these clumps.

To infer whether the identified clumps are stable and unstable against gravitational 
collapse, the virial mass analysis is employed. 
As mentioned above, we find six clumps associated with the filamentary structure 
and two of them are located at the waist of the bipolar nebula. 
The molecular line observations were reported toward four out of these six clumps.
We estimated virial mass (M$_{vir}$) and virial parameter (M$_{vir}$/$M_{clump}$) 
for these four clumps where line-width values were measured. 
The virial mass of a clump of radius R$_{c}$ (in pc) and line-width $\Delta V$ (in km s$^{-1}$) is defined 
as M$_{vir}$ ($M_\odot$)\,=\,k\,R$_{c}$\,$\Delta V^2$ \citep{maclaren88}, 
where the geometrical parameter, k\,=\,126, for a density profile $\rho$ $\propto$ 1/r$^2$. 
A virial parameter less than 1 suggests a clump prone to collapse, and greater than 1 is resistant to collapse. 
For these clumps, the virial parameter is less than 1, indicating unstable clumps against gravitational collapse. 

Following \citet{dewangan16}, we estimated thermal sound speed ($a_{s}$) 
and non-thermal velocity dispersion ($\sigma_{NT}$) of the two clumps where gas kinetic temperature (T$_{kin}$) values 
were reported \citep{wienen12}. \citet{wienen12} estimated T$_{kin}$ for clumps 1 and 2 to be 22.86 and 17.71 K, respectively. 
The sound speed $a_{s}$ (= $(k T_{kin}/\mu m_{H})^{1/2}$) is computed using T$_{kin}$ and 
$\mu$=2.37 (approximately 70\% H and 28\% He by mass). 
The non-thermal velocity dispersion is defined by:
\begin{equation}
\sigma_{\rm NT} = \sqrt{\frac{\Delta {V}^2}{8\ln 2}-\frac{k T_{kin}}{17 m_H}} = \sqrt{\frac{\Delta {V}^2}{8\ln 2}-\sigma_{\rm T}^{2}} ,
\label{sigmanonthermal}
\end{equation}
where $\Delta V$ is the measured line-width of the observed NH$_{3}$ spectra, 
$\sigma_{\rm T}$ (= $(k T_{kin}/17 m_H)^{1/2}$) is the thermal broadening for NH$_{3}$ at T$_{kin}$ \citep[e.g.][]{dunham11}, 
and $m_H$ is the mass of the hydrogen atom. 
We obtain $a_{s}$ and $\sigma_{NT}$ to be 0.28 and 1.42 for clump 1, respectively, 
while $a_{s}$ = 0.25 and $\sigma_{NT}$ = 1.44 for clump 2, respectively. 
These values give us the Mach Numbers ($\sigma_{\rm NT}$/$a_{s}$) to be 5 and 5.8 for clumps 1 and 2, respectively, 
indicating that these clumps are supersonic where non-thermal motions are dominated. 
The non-thermal motions could be explained by turbulence and outflows from YSOs.
\subsection{Velocity field}
\label{sec:coem} 
\begin{figure*}
\epsscale{0.5}
\plotone{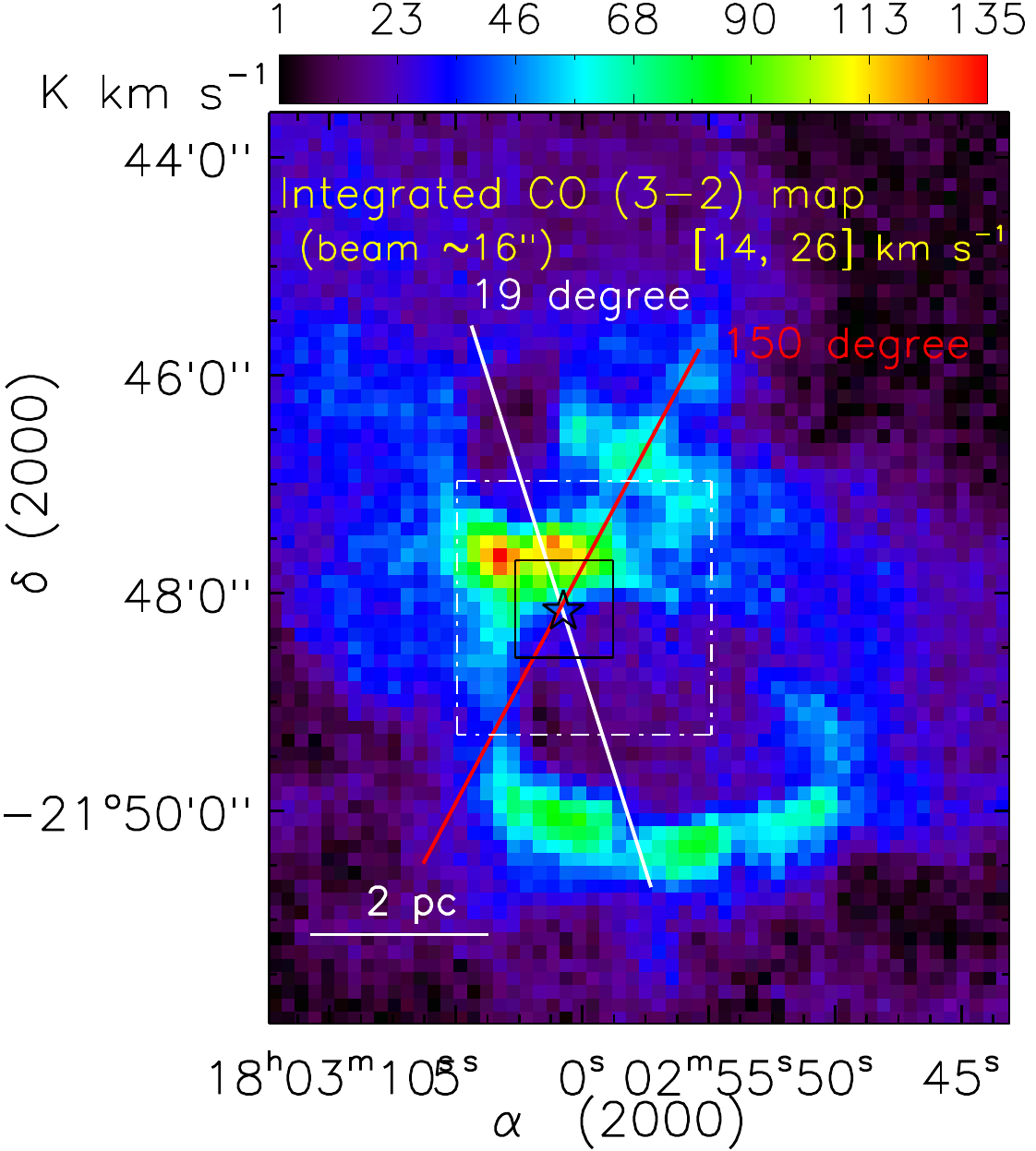}
\epsscale{1.1}
\plotone{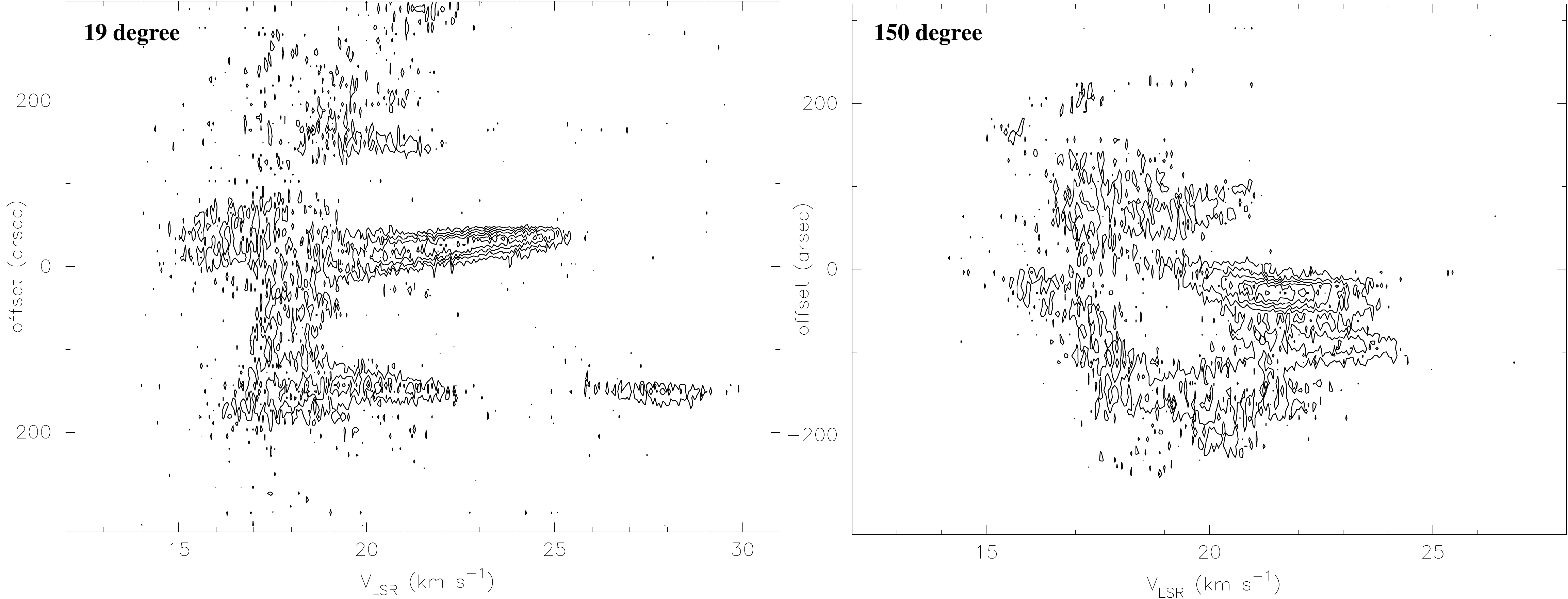}
\caption{\scriptsize \textbf{Top panel:} Integrated CO (3--2) emission map of the region 
around IRAS 17599$-$2148, similar as shown in Figure~\ref{fig2}b. 
Two solid lines with different position angles represent the axes, where the position-velocity diagrams are extracted. 
A star symbol indicates the position of the 6.7-GHz MME. 
The scale bar at the bottom-left corner corresponds to 2 pc (at a distance of 4.2 kpc). 
The dotted-dashed white box is shown as a zoomed-in view in Figure~\ref{fig6}a. 
The solid black box is shown as a zoomed-in view in Figure~\ref{fig8}a. 
\textbf{Bottom Left:} 
Position-velocity diagram along the axis with a position angle of 19$\degr$ (see top panel in Figure~\ref{fig5}).
\textbf{Bottom Right:} 
Position-velocity diagram along the axis with a position angle of 150$\degr$ (see top panel in Figure~\ref{fig5}). 
In both the bottom panels, the offset reference is the position of the 6.7-GHz MME. 
The position-velocity diagrams show a C-like and a ring-like structure of gas, indicating the expanding shell with an expanding gas 
velocity of $\sim$4.5 km s$^{-1}$.}
\label{fig5}
\end{figure*}

Previously, \citet{dewangan12} studied the G8.14+0.23 H\,{\sc ii} region and 
suggested the possibility of triggered star formation by the expansion of this H\,{\sc ii} region. 
In order to further examine the expansion of the H\,{\sc ii} region, we present a kinematic analysis of the molecular gas in the 
G8.14+0.23 H\,{\sc ii} region.  
In Figure~\ref{fig5}, the integrated CO intensity map and the position-velocity diagrams are presented. 
The position-velocity diagrams of the CO emission are shown with different position angles, tracing an C-like and a ring-like structure. 
Recently, \citet{arce11} studied the molecular gas distribution in the Perseus molecular cloud and compared a model 
of expanding bubbles in a turbulent medium. 
These authors predicted a C-like or ring-like structure for an expanding shell/bubble in the position-velocity 
diagrams (see Figure~5 in \citet{arce11}). 
In this work, the radio continuum maps suggest that the H\,{\sc ii} region associated with IRAS 17599$-$2148 
can be excited by a single source of radio spectral O6V--O6.5V. 
Hence, the structures seen in the position-velocity diagrams 
can be explained by the expanding H\,{\sc ii} region with an expansion velocity of the gas to be $\sim$4.5 km s$^{-1}$. 

Furthermore, the position-velocity diagrams show a noticeable velocity gradient toward the IRAS 17599$-$2148 region, 
indicating the presence of a molecular outflow(s). Due to coarse beam of CO (3-2) line data, 
it is indeed difficult task to explore the real driver(s) of the outflow(s). 
Figure~\ref{fig6}a shows the distribution of ionized and dust continuum emissions. 
It seems that the exciting source of the outflow could be driving source of the H\,{\sc ii} region or of the 6.7 GHz MME. 
Hence, high-resolution molecular line observations will be required to further explore the outflow activity in 
the IRAS 17599$-$2148 region. 

Due to lack of the JCMT molecular line observations toward the IRDC, the analysis presented in this 
work does not provide any information of the molecular gas kinematics toward the IRDC. 
\subsection{Small-scale environment toward IRAS 17599$-$2148}
\label{sec:small} 
In order to explore the driving source(s) of the 6.7 GHz MME and the radio continuum peaks, 
we examined the UKIDSS-GPS and GLIMPSE images toward IRAS 17599$-$2148 position.
We also utilized the ESO-VLT NIR images to probe the inner circumstellar environment of the driving sources. 
\subsubsection{Infrared counterpart of 6.7 GHz MME}
\begin{figure*}
\epsscale{0.53}
\plotone{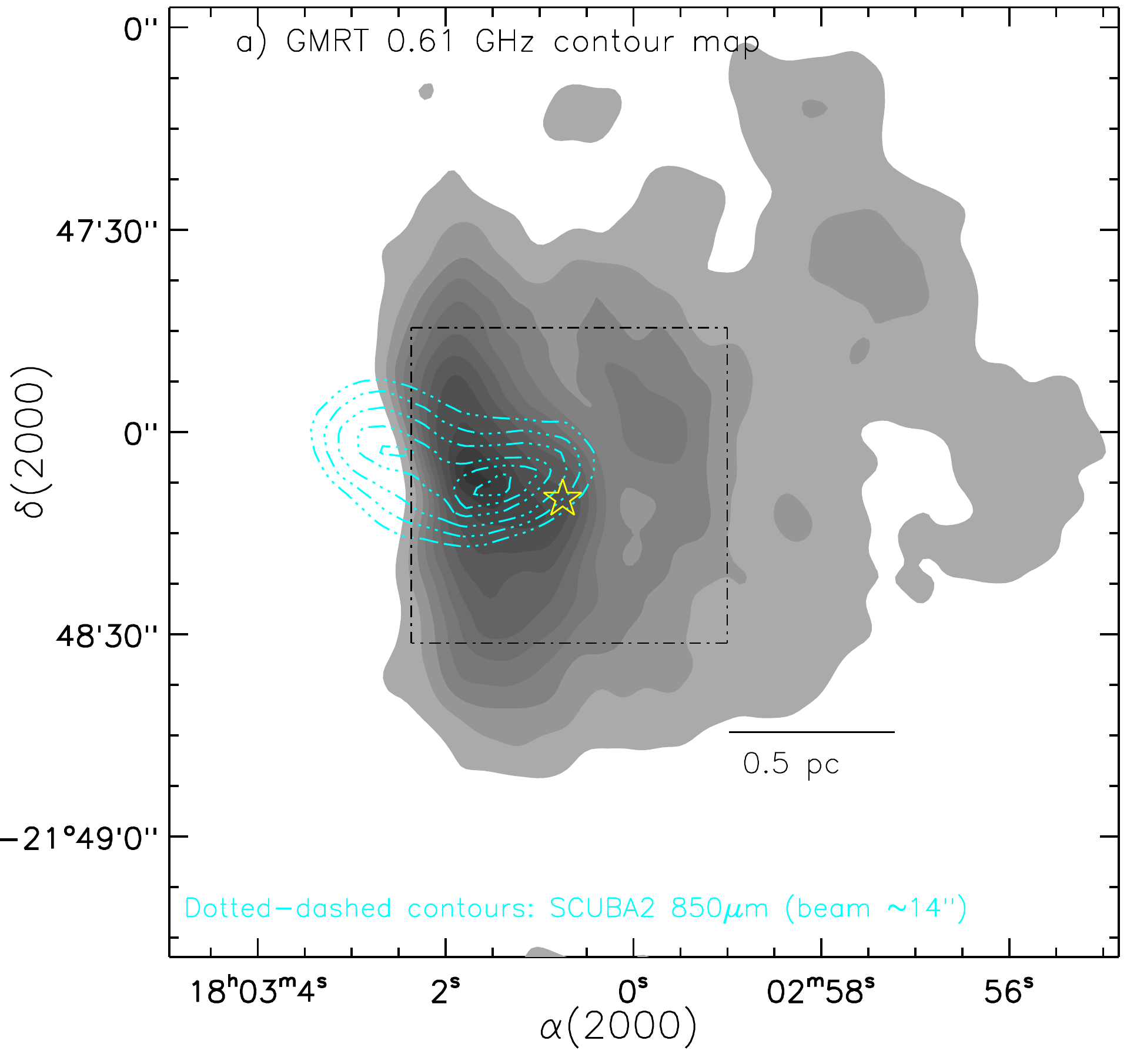}
\epsscale{0.6}
\plotone{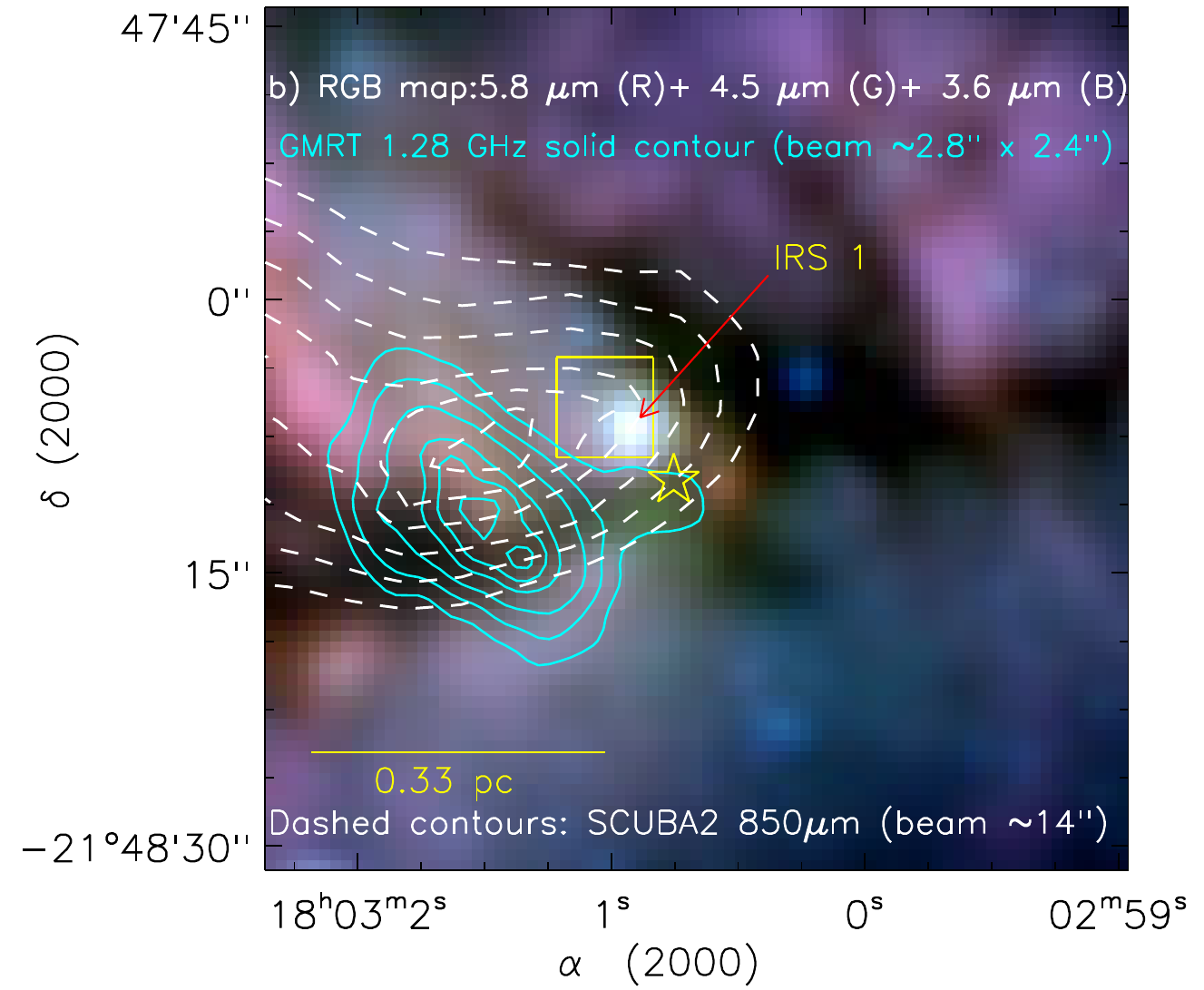}
\caption{\scriptsize a) The radio continuum contour map at 0.61 GHz is overplotted with 
contours of 850 $\mu$m emission (see dotted-dashed contours). 
The 0.61 GHz contours are shown with similar levels as shown in Figure~\ref{fig3}a. 
The dotted-dashed black box encompasses the area shown in Figure~\ref{fig6}b. 
The scale bar at the bottom-right corner corresponds to 0.5 pc (at a distance of 4.2 kpc). 
The figure shows the distribution of ionized and dust continuum emissions.
b) Three-color composite image ({\it red}, 5.8 $\mu$m; {\it green}, 4.5 $\mu$m; {\it blue}, 3.6 $\mu$m; in log scale) 
of the region around IRS~1. 
The 1.28 GHz emission is also overlaid by contours with levels of 60, 70, 80, 90, 95, and 99\% of the 
peak value (i.e. 0.0225 Jy/beam). IRS~1, previously reported by \citet{dewangan12}, 
appears the reddest and brightest in the {\it Spitzer} images, and is located 
close to the sub-mm emission peak as well as the Class~II 6.7 GHz MME position. 
The solid yellow box is shown as a zoomed-in view in Figures~\ref{fig7}a,~\ref{fig7}b, and~\ref{fig7}c.
The scale bar at the bottom-left corner corresponds to 0.33 pc (at a distance of 4.2 kpc). 
In both the panels, the 850 $\mu$m emission is also overlaid by contours 
with levels of 60, 70, 80, 90, 95, and 99\% of the peak value (i.e., 15.5 mJy/arcsec$^{2}$). 
In both the panels, the position of the 6.7-GHz MME is marked with a star symbol.} 
\label{fig6}
\end{figure*}

Figure~\ref{fig6}b shows a three-color composite image made using the GLIMPSE images 
(5.8 $\mu$m in red; 4.5 $\mu$m in green; 3.6 $\mu$m in blue), 
which is also overlaid with the 1.28 GHz and 850 $\mu$m emission. 
A point-like source appears the reddest and brightest in the {\it Spitzer} images, and is located very close to 
the sub-mm emission peak (a projected linear separation of $\sim$8$\arcsec$) and the 6.7-GHz MME (a projected linear separation of $\sim$4$\arcsec$). 
The source was previously designated as IRS~1 and was 
characterized as a candidate MYSO (stellar mass $\sim$10$\pm$2~M$_{\odot}$) based on its SED modeling 
from the NIR to sub-mm data \citep[see Figure~11 in][]{dewangan12}. 
IRS~1 is barely observed in the GPS H-band image (not shown here), however, it is 
seen as a single object in the GPS K-band image (resolution $\sim$0.8$\arcsec$; not shown here) 
and the GLIMPSE 3.6--8.0 $\mu$m images (resolution $\sim$2$\arcsec$) (see Figure~\ref{fig6}b). 
IRS~1 is an embedded source located very close to the peak of the 6.7 GHz MME and is associated with a clump having the highest column density. 
Additionally, considering IRS~1 as a single bright source in the GLIMSE images, we consider the IRS~1 source as an infrared counterpart (IRc) of the 6.7-GHz MME.  
At the sensitivity of the observed GMRT radio map (3$\sigma$ $\sim$1.14 mJy beam$^{-1}$), 
IRS~1 is situated at a projected linear separation of $\sim$10$\arcsec$ from the radio 
peaks seen in the 1.28 GHz map, which suggests that IRS~1 is not associated with any radio continuum peak 
(or an ultracompact (UC) H\,{\sc ii} region; see Figure~\ref{fig6}b). 
Hence, IRS~1 cannot be the powering source of the H\,{\sc ii} region associated with the IRAS 17599$-$2148 region.  
As mentioned before, the 6.7-GHz MME is a reliable tracer of MYSO \citep[e.g.][]{walsh98,minier01,urquhart13}. 
All together, IRS~1 could be a genuine massive protostar candidate (mass $\sim$10$\pm$2~M$_{\odot}$) in a very early evolutionary stage, 
before the onset of an UCH\,{\sc ii} region. 
Based on the presence of two compact radio continuum sources (with the radio spectral class of O-type stars) and the MME, IRAS 17599$-$2148 region appears to be a very similar system like W42 
\citep[see Figure~2 in][]{dewangan15b}. 

\begin{figure*}
\epsscale{1.1}
\plotone{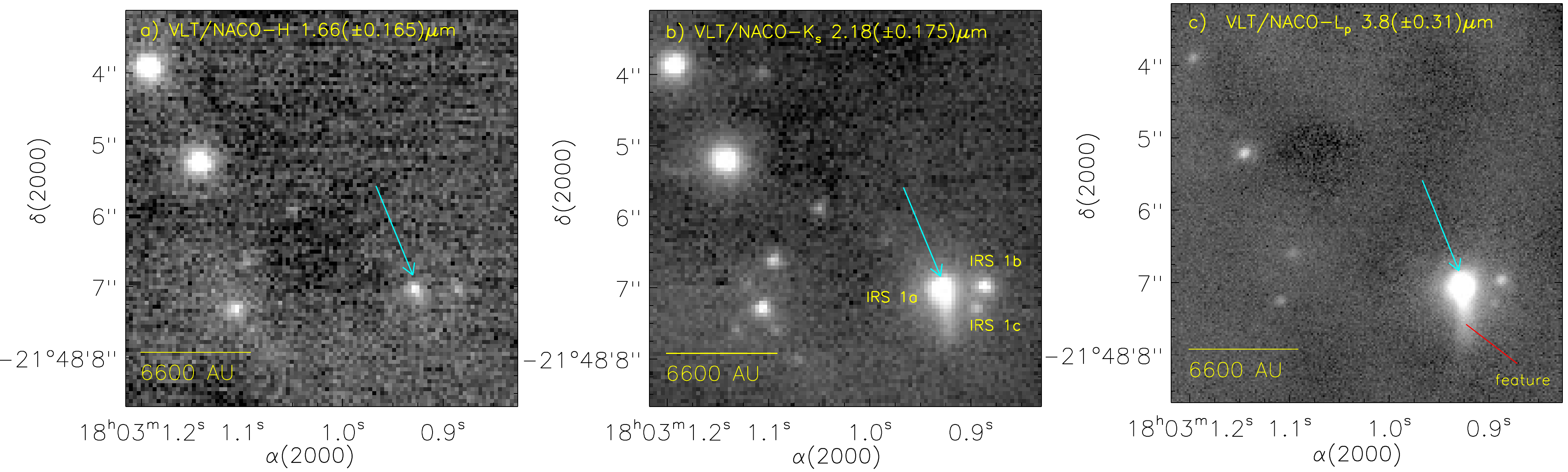}
\caption{\scriptsize The VLT/NACO adaptive-optics H, K$_{s}$, and L$^{\prime}$ images of the region around IRS~1 (in log scale). 
The selected area of the VLT/ NACO images is shown by a solid box in Figure~\ref{fig6}b. 
In each panel, the position of a resolved point-like source ($\alpha_{2000}$ = 18$^{h}$03$^{m}$00$^{s}$.93, $\delta_{2000}$ = $-$21$\degr$48$\arcmin$07$\arcsec$.05) is highlighted by an arrow. In each panel, the scale bar at the 
bottom-left corner corresponds to 6600 AU (at a distance of 4.2 kpc). 
The NACO images have resolved IRS~1 into at least three point-like sources (labeled as IRS 1a, 1b, and 1c) and one of 
them is associated with a diffuse emission feature within a scale of 4200 AU.} 
\label{fig7}
\end{figure*}

In order to study the small scale environment of IRS~1, we present the ESO-VLT archival NIR images. 
In Figure~\ref{fig7}, we show the VLT/NACO adaptive-optics images of IRS~1 in H, K$_{s}$, and L$^{\prime}$ bands. 
The VLT/NACO H-band image detects a faint point-like source, while K$_{s}$ and L$^{\prime}$ images resolve the IRS~1 
into at least three point-like sources (i.e. designated as IRS~1a, IRS~1b, and IRS~1c; see Figure~\ref{fig7}b) within a scale of $\sim$4200 AU 
and one of the sources (i.e. IRS~1a) is associated with the diffuse emission (see Figures~\ref{fig7}b and~\ref{fig7}c). 
We find these sources in the separation range from $\sim$0\farcs33 (1390 AU) to $\sim$0\farcs57 (2400 AU). 
\citet{devilliers15} summarized that the 6.7 GHz masers switch on after the onset of the outflow. 
The diffuse emission seen below $\sim$4200 AU scale with IRS~1a might be linked with the outflow. 
In general, the NIR emission is attributed to scattered light escaping through the outflow cavity \citep[e.g.][]{zhang11}. 
Hence, the diffuse emission feature could be a jet, which appears as a narrow emission feature and 
may be located inside an outflow cavity as traced in the L$^{\prime}$ image. 
Additionally, \citet{urquhart13} suggested the exclusive association of Class~II 6.7 GHz methanol masers with MYSOs. 
Hence, IRS~1a could be the main massive protostar among other point-like sources in the IRS~1 system. 

Additionally, the multiplicity of point-like sources has been observed around IRS~1 within a scale of 4200~AU. 
Massive stars are generally found in binary and multiple systems \citep{duchene13}. 
For example, in M8 and W42 star-forming regions, the well characterized ``O" stars are resolved into multiple sources using the 
high-resolution NIR images \citep{goto06,dewangan15b}. 
Around IRS~1, the multiple sources are found in the separation range from $\sim$0\farcs33 (1390 AU) to $\sim$0\farcs57 (2400 AU). 
Due to the coarse beam of the JCMT-SCUBA-2, the 850 $\mu$m map is unable to resolve the dense core associated with IRS~1 source. 
Based on the separation of the sources in IRS~1 multiple system, there is possibility of the core fragmentation that 
may explain the origin of multiple systems \citep[see][and references therein]{tobin16}. 

Based on these indicative results, 
further detailed studies of IRS~1 are encouraged using high-resolution interferometric observations at longer wavelength.
\subsubsection{Infrared counterparts of compact radio peaks}
\begin{figure*}
\epsscale{1.0}
\plotone{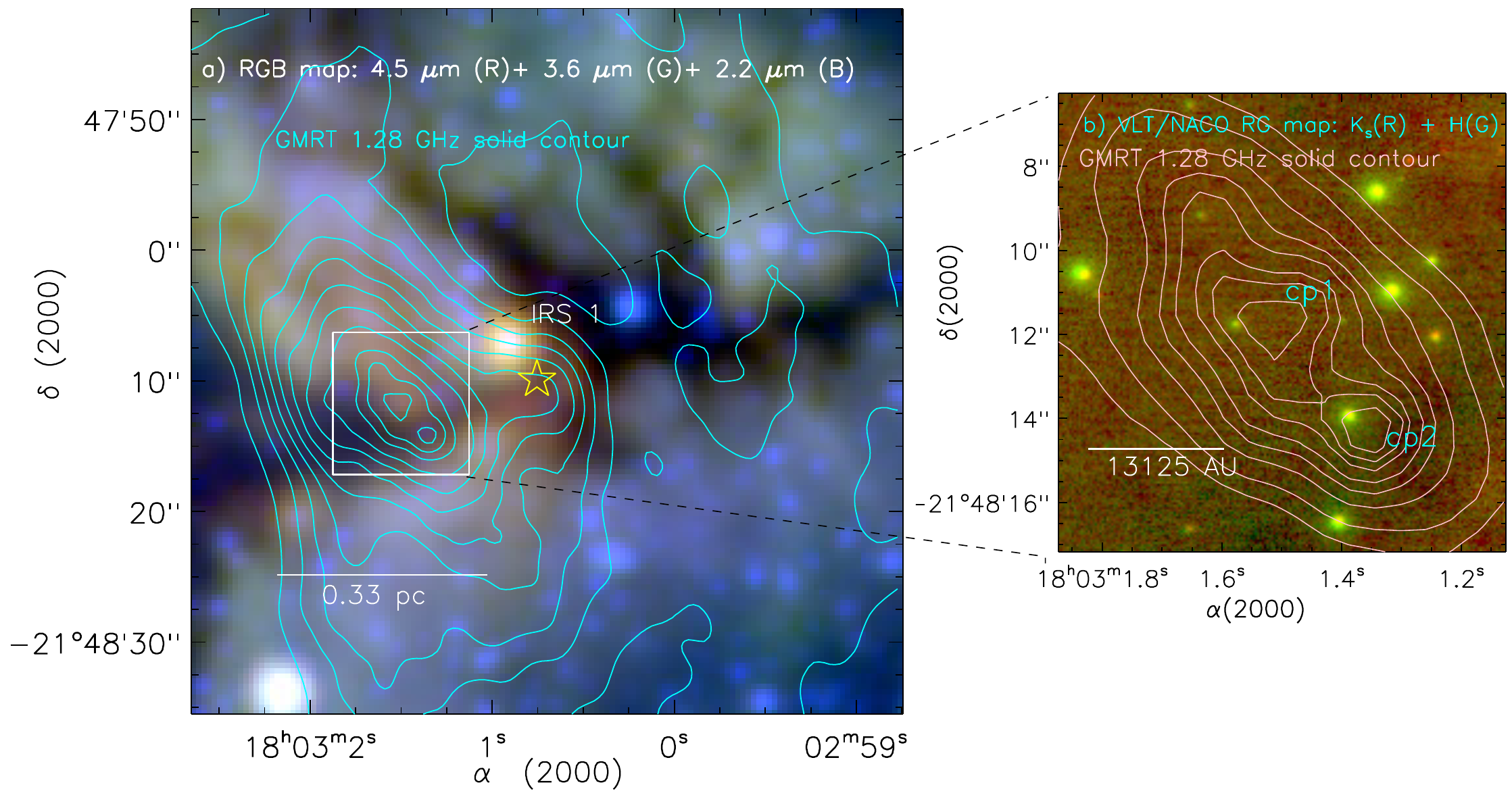}
\caption{\scriptsize a) 
Three-color composite image ({\it red}, 4.5 $\mu$m; {\it green}, 3.6 $\mu$m; {\it blue}, 2.2 $\mu$m; in log scale) 
of the region around IRS~1. The 1.28 GHz emission is also overlaid by contours 
with similar levels as shown in Figure~\ref{fig3}b. 
The solid white box encompasses the area shown in Figure~\ref{fig8}b. A star symbol indicates the position of the 6.7-GHz MME. 
The scale bar at the bottom-left corner corresponds to 0.33 pc (at a distance of 4.2 kpc). 
b) The VLT/NACO two-color composite image ({\it green}, H-band; {\it red}, K$_{s}$-band; in log scale) of the region 
around the radio compact peaks. 
The selected area of the VLT/NACO images is shown by a box in Figure~\ref{fig8}a, tracing the small-scale environment. 
The 1.28 GHz radio contours are also overlaid with levels of 80, 85, 90, 93, 95, 97, 98, and 99\% of the peak 
value (i.e. 0.0225 Jy/beam). The scale bar at the bottom-left corner corresponds to 13125 AU (at a distance of 4.2 kpc). 
In the VLT/NACO images, a point-like source (without any diffuse emission) is seen toward each radio peak. }
\label{fig8}
\end{figure*}

To trace the IRcs of two compact radio peaks (cp1 and cp2), Figure~\ref{fig8}a shows a three-color composite image using 
4.5 $\mu$m (red), 3.6 $\mu$m (green), and 2.2 $\mu$m (blue), which is also overlaid with the 1.28 GHz emission. 
The source seen close to the peak cp1 is barely observed in the GPS K-band image. 
The source seen close to the peak cp2 is also observed in the GPS K-band image 
(position and JHK photometry: $\alpha_{2000}$ =18:03:01.38, $\delta_{2000}$ = $-$21:48:13.96; m$_{J}$ = 18.22$\pm$0.07 mag; 
m$_{H}$ = 17.39$\pm$0.08 mag; m$_{K}$ = 17.79$\pm$0.3 mag). 
In Figure~\ref{fig8}b, we present the VLT/NACO adaptive-optics NIR images (H and K$_{s}$; below 8000 AU scale) 
toward the radio peaks (i.e. cp1 and cp2). 
A point-like source is seen toward each compact radio peak in the VLT/NACO images. 
There are no nebular features seen toward these sources in the NACO images. 

High-resolution spectroscopic study will be very helpful to confirm the ionizing sources of the radio peaks. 
\subsection{Young stellar objects}
\subsubsection{Study of identification of YSOs}
\label{subsec:phot1} 
\begin{figure*}
\epsscale{0.98}
\plotone{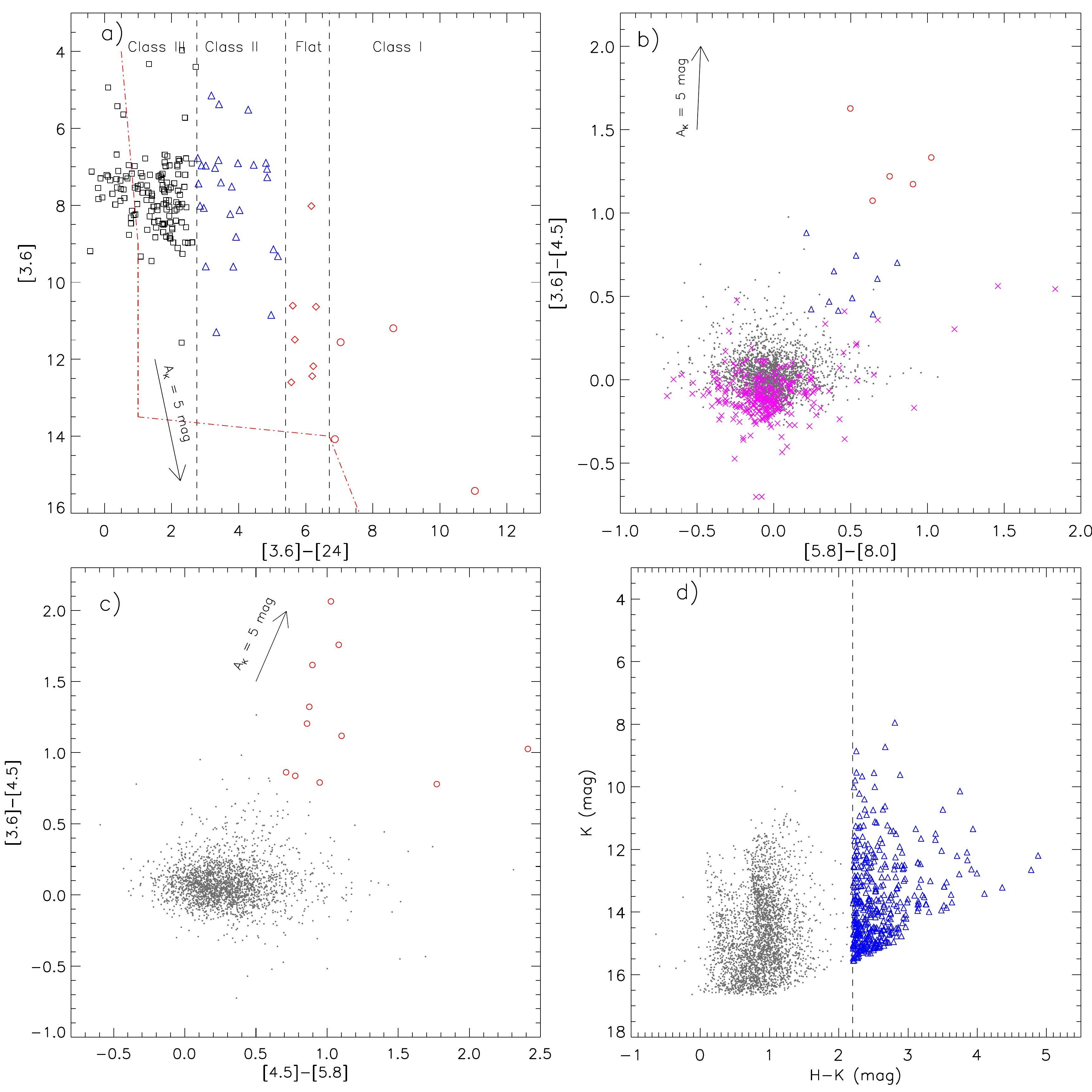}
\caption{\scriptsize Selection of embedded young stellar population within the region probed in this paper (see Figure~\ref{fig2}). 
a) Color-magnitude diagram ([3.6] $-$ [24] vs [3.6]) showing the YSOs belonging to different evolutionary stages (see dashed lines).
These sources are detected in the GLIMPSE 3.6 $\mu$m and the MIPSGAL 24 $\mu$m bands (see the text for more details). 
The YSOs and contaminated candidates (galaxies and disk-less stars) are separated by a dotted-dashed 
curve \citep[see][for more details]{rebull11}. 
The Flat-spectrum and Class~III sources are designated by ``$\Diamond$'' and ``$\Box$'' symbols, respectively; 
b) Color-color diagram ([3.6]$-$[4.5] vs. [5.8]$-$[8.0]) of sources detected in the four GLIMPSE bands. 
The ``$\times$'' symbol shows the PAH-emission-contaminated apertures (see the text); 
c) Color-color diagram ([3.6]$-$[4.5] vs. [4.5]$-$[5.8]) using the first three GLIMPSE band detections (except 8.0 $\mu$m image); 
d) Color-magnitude diagram (H$-$K/K) of the sources observed in the NIR bands. 
In all the panels, Class~I and Class~II YSOs are marked by red circles and open blue triangles, respectively. 
In the last three panels, the dots in gray color refer the stars with only photospheric emissions. 
In the NIR H$-$K/K space, we have shown only 3001 out of 33984 stars with photospheric emissions. 
In the first three panels, the extinction vectors are drawn using the average extinction laws from \citet{flaherty07}. 
The positions of all the identified YSOs are shown in Figure~\ref{fig10}a.}
\label{fig9}
\end{figure*}

The study of YSO populations is a powerful tool to depict the picture of ongoing star formation activity in a given star-forming region.
In this section, we identify and classify the YSO populations using their infrared excess for a wider field of view around IRAS 17599$-$2148. 
Previously, \citet{dewangan12} also presented the photometric analysis of point-like sources in the IRAS 17599$-$2148 region, 
which was restricted only near the bipolar nebula. 
In the following, we give a brief description of YSOs identification and classification schemes adopted in this work.\\\\ 
1. In this scheme, the GLIMPSE-MIPSGAL color-magnitude diagram ([3.6]$-$[24]/[3.6]) is used to identify the different stages of 
YSOs \citep{guieu10,rebull11,dewangan15b}, where the sources have detections in the MIPSGAL 24 $\mu$m and GLIMPSE 3.6 $\mu$m bands.
 The 24 $\mu$m image is saturated near the IRAS position, therefore this scheme is not used to trace YSOs near the IRAS position (see Figure~\ref{fig1}a).
 Following the conditions given in \citet{guieu10}, the boundary of different stages of YSOs is also highlighted in Figure~\ref{fig9}a. 
 In our selected region, we find 170 sources that have detections in both the 3.6 and 24 $\mu$m bands.  
 We identify 39 YSOs (4 Class I; 7 Flat-spectrum; 28 Class~II) and 131 Class~III sources. 
 Furthermore, Figure~\ref{fig9}a exhibits the boundary of possible contaminants 
 (i.e. galaxies and disk-less stars) \citep[also see Figure~10 in][]{rebull11}. 
We also find that our identified YSO populations are free from the contaminants. 
We also identify an embedded source in the IRDC, which is detected at wavelengths longer than 3.6 $\mu$m and 
is designated as MG008.2845+00.1664 in the MIPSGAL photometry catalog \citep{gutermuth15}. 
The source has 4.5--24 $\mu$m bands photometry and is a YSO candidate with a color (m$_{8.0}$ - m$_{24}$) of 3.54 mag.\\
 
2. In this scheme, the GLIMPSE color-color diagram ([3.6]$-$[4.5] vs [5.8]$-$[8.0]) of sources is employed 
that have detections in all four GLIMPSE bands. 
Following the \citet{gutermuth09} schemes, YSOs and various possible contaminants (e.g. broad-line active galactic nuclei (AGNs), 
PAH-emitting galaxies, shocked emission blobs/knots, and PAH-emission-contaminated apertures) are identified in our selected region.
The possible contaminants are also removed from the selected YSOs. 
Furthermore, these YSOs are classified into different evolutionary stages based on their 
slopes of the {\it Spitzer}-GLIMPSE SED ($\alpha_{3.6-8.0}$) computed from 3.6 to 8.0 $\mu$m 
(i.e. Class~I ($\alpha_{3.6-8.0} > -0.3$), Class~II ($-0.3 > \alpha_{3.6-8.0} > -1.6$), 
and Class~III ($-1.6> \alpha_{3.6-8.0} > -2.56$)) \citep[e.g.,][]{lada06}. 
The GLIMPSE color-color diagram ([3.6]$-$[4.5] vs [5.8]$-$[8.0]) is shown in Figure~\ref{fig9}b. 
More details on the YSO classifications can be found in the work of \citet[][and references therein]{dewangan11}. 
This scheme leads 15 YSOs (5 Class I; 10 Class~II), 2282 photospheres, and 288 contaminants.\\ 

3. In this scheme, the color-color space ([4.5]$-$[5.8] vs [3.6]$-$[4.5]) of sources is used 
that have detections in the first three GLIMPSE bands (except 8.0 $\mu$m band). 
The color conditions, [4.5]$-$[5.8] $\ge$ 0.7 and [3.6]$-$[4.5] $\ge$ 0.7, are adopted to select protostars, 
as given in \citet{hartmann05} and \citet{getman07}. 11 protostars are found in our selected region (see Figure~\ref{fig9}c). \\ 

4. In this scheme, the color-magnitude space (H$-$K/K) is adopted for selecting additional YSOs. 
To find a color condition, we explored the color-magnitude space of the nearby control field 
(size $\sim$5$\arcmin$  $\times$ 5$\arcmin$; central coordinates: $\alpha_{J2000}$ = 18$^{h}$02$^{m}$20$^{s}$.7, 
$\delta_{J2000}$ = $-$21$\degr$50$\arcmin$31$\arcsec$.1) and selected a color H$-$K value (i.e. $\sim$2.2) that 
separates large H$-$K excess sources from the rest of the population. 
Considering this color H$-$K cut-off criterion, 409 embedded YSOs are identified in our 
selected region (see Figure~\ref{fig9}d).\\

Finally, all these schemes give us a total of 474 YSOs in our selected region 
(as shown in Figure~\ref{fig1}). The positions of all these YSOs are shown in Figure~\ref{fig10}a.
Several embedded sources are detected toward the elongated filamentary structure. 
%\newpage
\subsubsection{Study of distribution of YSOs}
\label{subsec:surfden} 
\begin{figure*}
\epsscale{0.9}
\plotone{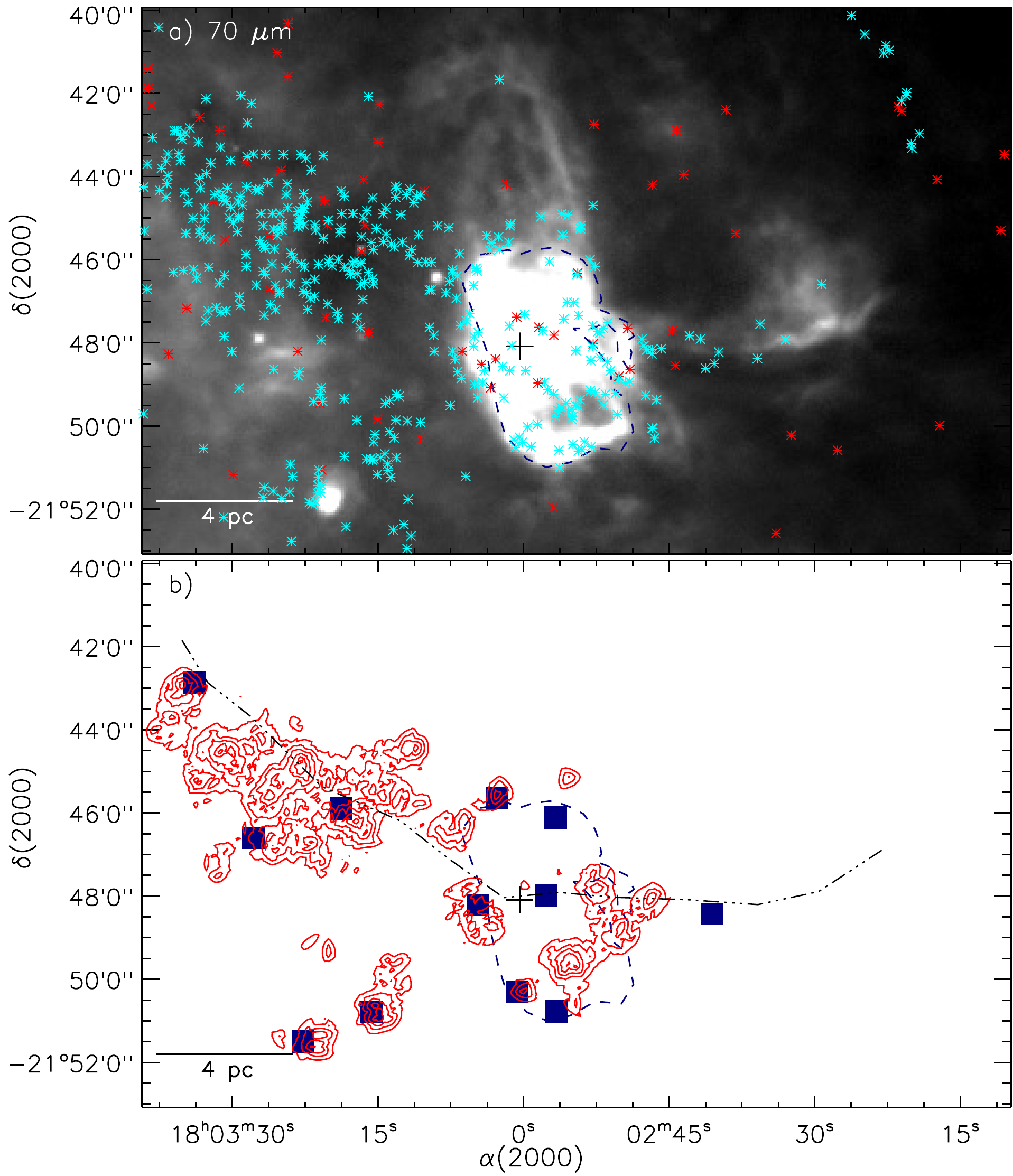}
\caption{\scriptsize The spatial distribution of YSOs identified within the region probed in this paper using the NIR and MIR data. 
a) The asterisks represent the identified YSOs, which are overlaid on the 70 $\mu$m image. 
The YSOs selected using the GLIMPSE and MIPSGAL bands (see the first three panels in Figure~\ref{fig9}) are shown by red color, 
whereas the cyan asterisks represent the YSOs identified using the NIR bands (see Figure~\ref{fig9}d). 
b) Figure shows the surface density contours (in red) of all the identified YSOs. 
The contours are shown at 4, 6, 9, 15, and 25 YSOs/pc$^{2}$, from the outer to the inner side. 
The filled squares represent the positions of the identified {\it Herschel} clumps (see Table~\ref{tab1}). 
A tight spatial association between YSOs clusters and clumps is evident. 
A bipolar nebula is also highlighted in both the panels similar to the one shown in Figure~\ref{fig4}. 
An elongated filamentary structure is highlighted by a dashed-dotted curve. 
The position of IRAS 17599$-$2148 (+) is shown in both the panels. 
In both the panels, the scale bar at the bottom-left corner corresponds to 4 pc (at a distance of 4.2 kpc). 
Star formation activities are revealed toward the IRDC and bipolar nebula (see text for details).}
\label{fig10}
\end{figure*}

The spatial distribution of young stellar populations is often examined 
using their surface density analysis \citep[e.g.][]{gutermuth09,bressert10}, which allow us to infer the young stellar clusters.
Previously, using the nearest-neighbour (NN) technique, 
\citet{dewangan12} also obtained the surface density map of YSOs in the IRAS 17599$-$2148 region, 
which was studied only near the bipolar nebula. 
The surface density map can be generated by dividing the selected field using a regular grid and computing 
the surface density of YSOs at each grid point. 
The surface number density at the {\it j$^{th}$} grid point is defined 
by $\rho_{j} = (n-1)/A_{j}$ \citep[e.g.][]{casertano85}, where $A_{j}$ represents 
the surface area defined by the radial distance to the $n$ = 6 NN.
Following this procedure, we generated the surface density map of all the selected 474 YSOs, 
using a 5$\arcsec$ grid and 6 NN at a distance of 4.2 kpc. 
In Figure~\ref{fig10}b, the resultant surface density contours of YSOs is presented. 
The contour levels are drawn at 4, 6, 9, 15, and 25 YSOs/pc$^{2}$, 
increasing from the outer to the inner regions. 
The YSOs clusters are mainly seen toward the IRDC and the bipolar nebula (see Figure~\ref{fig10}b). 
The star formation activity is found toward all the condensations as 
traced in the {\it Herschel} column density map.

We also obtained the clustered populations from distributed sources using a statistical analysis of YSOs. 
We estimated an empirical cumulative distribution (ECD) of YSOs as a function of NN 
distance \citep[see][for more details]{chavarria08,gutermuth09,dewangan11}.
In the ECD analysis, a cutoff length (also referred as the distance of inflection d$_{c}$) is selected 
for delineating the low-density/distributed populations. 
For the present case, we selected a cutoff distance of d$_{c}$ $\sim$44$\arcsec$ (0.9 pc at a distance of 4.2 kpc). 
The ECD analysis provided a clustered fraction of about 72\% YSOs (i.e. 344 from a total of 474 YSOs). 
\section{Discussion}
\label{sec:disc}
On a wider field of view around IRAS 17599$-$2148, an elongated filamentary structure (extension $\sim$21 pc) 
has been traced in the {\it Herschel} continuum images. 
In Section~\ref{subsec:u16}, we mentioned about the presence a continuous velocity structure in the direction of IRAS 17599$-$2148 and 
presented evidence of the existence of the elongated filamentary structure. 
Considering the molecular and ionized gas velocities, the elongated filamentary structure harbours the IRDC, and 
a bipolar nebula associated with the H\,{\sc ii} region IRAS 17599$-$2148/G8.14+0.23. 
The ATLASGAL sources at 870 $\mu$m (i.e., a chain of dense clumps/cores) are exclusively found toward the IRDC and the bipolar nebula, tracing dense regions qualitatively.
In Section~\ref{subsec:temp}, using {\it Herschel} data analysis, the physical conditions (column density and temperature) have been inferred in the region 
around IRAS 17599$-$2148. 
In the {\it Herschel} temperature map, the IRDC is traced with cold gas (without any radio detection), 
while the bipolar nebula is seen with warmer gas, indicating a temperature gradient. 
We found higher temperatures toward the waist of the bipolar nebula which is part of the H\,{\sc ii} region. 
This implies that the waist of the bipolar nebula is heated directly by massive stars that are embedded in dense gas as traced by NH$_{3}$ emission. 
Based on the temperature distribution in the {\it Herschel} temperature map, 
the bipolar nebula is clearly distinguished from the IRDC. 
This immediately suggests the presence of two distinct environments within the elongated filamentary structure. 
In our selected region, the {\it Herschel} column density map reveals twelve clumps. 
We find at least six massive clumps (IDs: 1--5 and 10; M$_{clump}$ $\sim$777 -- 7024 M$_{\odot}$) 
associated with the filamentary structure and two of them (IDs:1--2) are located at the waist of the bipolar nebula.
Four additional clumps (IDs: 6--8 and 12) are also identified at the edges of the bipolar nebula. 
All the twelve clumps have higher self-gravitating pressure values (i.e. 2.1--29.2 $\times$ 10$^{-10}$ dynes cm$^{-2}$) 
and smaller $t_{ff,sph}$ values (i.e. 0.4--0.77 Myr), indicating the signature of the early phase of star formation within the clumps. 
Based on availability of the line parameters toward four clumps (IDs:1--4), 
the virial parameters (M$_{vir}$/$M_{clump}$) for these clumps are less than 1, suggesting these clumps are prone to collapse.
Based on availability of T$_{kin}$, two (IDs:1--2) out of these four clumps have higher Mach numbers ($\sigma_{\rm NT}$/$a_{s}$)  (i.e.  5 and 5.8), indicating 
 that the clumps are dominated by non-thermal motions (such as outflows from YSOs). These clumps appear supersonic. 
 All these results favour that the fragmentation occurred within the filaments, 
 which leads several clumps with higher column densities. Indirectly, these clumps show the signatures of ongoing star formation activities. 
 
 In the era of ATLASGAL and {\it Herschel} surveys, the identification of elongated filaments and the formation of clumps 
in these filaments have received much attention \citep[e.g.][and references therein]{schneider12,ragan14,contreras16,li16}. 
The large-scale filaments are likely to be unstable to radial collapse and fragmentation. 
The spacing of the clumps in the filamentary cloud is often explained by the ``sausage" instability produces 
during the gravitational collapse of a cylinder \citep{chandrasekhar53,nagasawa87}. 
More recently, \citet{contreras16} studied fragmentation in five filamentary molecular clouds 
using data from the ATLASGAL 870 $\mu$m, {\it Spitzer}, and Millimetre Astronomy Legacy Team
90 GHz (MALT90) surveys and also suggested the observed separation of the clumps via the ``sausage" instability theory. 
In this work, the presence of clumps (or a chain of dense clumps/cores) in the elongated filamentary structure could be explained by this instability theory. 
However, in order to validate the theoretical predictions, high-resolution molecular line data toward the elongated structure are required. \\

The distribution of YSOs clusters traces the star formation activities. 
In Section~\ref{subsec:surfden}, the YSOs clusters are found toward the clumps in the elongated filamentary structure.  
The association of clumps and YSOs clusters is evident, illustrating further confirmation of star formation within the clumps. 
Additionally, the spatial distribution of YSOs clusters is also seen at the edges of the bipolar nebula. 
Considering the spatial locations of the clusters, star formation appears more intense toward the IRDC.
Note that the IRDC contains many deeply embedded YSOs without any ionized emission.
The ionized emission is exclusively associated with the most massive clump, which is located at the waist of the bipolar nebula.
A bipolar appearance of IRAS 17599$-$2148 could be explained due to the ionizing feedback from the O6.5-O6 type star. 
The position-velocity analysis of the gas kinematics of CO(3--2) suggests 
the signature of an expanding H\,{\sc ii} region with an expansion velocity of the gas to be $\sim$4.5 km s$^{-1}$.
\citet{dewangan12} estimated the dynamical or expansion age of the H\,{\sc ii} region to be $\sim$ 1.6 Myr.
The average ages of Class~I and Class~II YSOs are estimated to be $\sim$0.44 Myr and $\sim$1--3 Myr \citep{evans09}, respectively. 
Hence, star formation in the clumps associated with the edges of the bipolar nebula could be influenced by the H\,{\sc ii} region. 
The H\,{\sc ii} region is spatially situated far away from the IRDC, therefore the star formation toward the IRDC is unlikely 
influenced by the H\,{\sc ii} region. 
This argument is also supported by the lower value of pressure of the H\,{\sc ii} region 
(i.e. $P_{HII}= \mu m_{H} c_{s}^2\, \left(\sqrt{3N_{uv}\over 4\pi\,\alpha_{B}\, D_{s}^3}\right)$ $\approx$ 8--5 $\times$ 10$^{-11}$ dynes\, cm$^{-2}$; 
where, $\mu$ = 2.37, c$_{s}$ is the sound speed of the photo-ionized gas (= 10 km s$^{-1}$), 
and $\alpha_{B}$ is the radiative recombination coefficient (= 2.6 $\times$ 10$^{-13}$ cm$^{3}$ s$^{-1}$)) 
driven by an O6.5-O6 type star at projected distances (D$_{s}$) of 8--11 pc, which is very close to P$_{MC}$. 
The onset stellar cluster formation associated with the IRDC appears to be spontaneous.
Combining these results, we notice that the different star formation processes have taken place in the bipolar nebula and the IRDC. 
The star formation associated with the H\,{\sc ii} region is likely more evolved compared to the IRDC.
\section{Summary and Conclusions}
\label{sec:conc}
The present work deals with a multi-scale and multi-wavelength analysis around IRAS 17599$-$2148, 
using new observations along with publicly available archival datasets. 
The goal of this paper is to understand the physical environment and star formation processes around IRAS 17599$-$2148 
on the smaller and larger scales. In the following, the important results of this work are provided.\\
$\bullet$ On a larger scale, an elongated filamentary structure is evident in the {\it Herschel} images, 
which hosts the IRDC and the H\,{\sc ii} region IRAS 17599$-$2148/G8.14+0.23. 
IRAS 17599$-$2148 has a bipolar appearance at wavelengths longer than 2 $\mu$m. Radio emission is not detected toward the IRDC. \\  
$\bullet$ High-resolution GMRT radio continuum emissions at 0.61 GHz (beam size $\sim$5\farcs6 $\times$ 5\farcs2)
and 1.28 GHz (beam size $\sim$2\farcs8 $\times$ 2\farcs4) show an extended radio emission, 
which is located toward the waist of the bipolar nebula and can be excited by a single star of radio spectral type, O6.5--O6.\\ 
$\bullet$ The 1.28 GHz map reveals two radio continuum peaks (cp1 and cp2) toward the single peak seen in the 0.61 GHz map. The radio peaks, cp1 and cp2 corresponds to 
a single ionizing star of radio spectral type, O8V--O8.5V and O9V--O9.5V, respectively.\\  
$\bullet$ Within a 18$\arcsec$ of the IRAS 17599$-$2148 position, two radio O-spectral type sources and a 6.7 GHz MME (without any radio peak) 
are traced, indicating the presence of different early evolutionary stages of massive star formation. \\
$\bullet$ IRS~1 has been identified as an IRc of the 6.7-GHz MME and is a massive protostar candidate. 
The GMRT radio continuum map at 1.28 GHz (1$\sigma$ $\sim$0.38 mJy beam$^{-1}$) does not trace any radio peak toward IRS~1.\\ 
$\bullet$ The inner circumstellar environment of IRS~1 is mapped using the VLT/NACO adaptive-optics 
K$_{s}$ and L$^{\prime}$ observations at resolutions~$\sim$0\farcs2 and $\sim$0\farcs1, respectively. 
Within a scale of 4200~AU, IRS~1 has been resolved into at least three point-like sources and one of them is associated with diffuse emission, 
which could be a collimated infrared jet.\\ 
$\bullet$ Multi-scale and multi-wavelength data suggest that IRS~1 could be a genuine massive protostar candidate in a very early evolutionary stage, 
prior to an UCH\,{\sc ii} phase. Further detailed studies of IRS~1 are encouraged using high-resolution interferometric observations 
at longer wavelength.\\ 
$\bullet$ The {\it Herschel} column density map traces twelve clumps and their masses vary between 410 M$_{\odot}$ 
and 7024 M$_{\odot}$. 
In the selected region around IRAS 17599$-$2148, these clumps are identified with a range of 
temperature and column density of about 16--39~K and 0.6--11~$\times$~10$^{22}$ cm$^{-2}$ (A$_{V}$ $\sim$7--117 mag), respectively.\\
$\bullet$ The elongated filamentary structure is associated with at least six massive condensations, indicating the fragmentation of the cloud. 
 This argument is also supported by virial analysis of clumps and their higher values of self-gravitating pressure. 
 The most massive condensation (mass $\sim$7024 M$_{\odot}$) hosts two radio continuum peaks and the 6.7 GHz MME, 
which is located at the waist of the bipolar nebula. At least four condensations are also found at the edges of the bipolar nebula.\\ 
$\bullet$ In the {\it Herschel} temperature map, the IRDC is found with cold gas (and the absence of any radio continuum emission), 
while the bipolar nebula is associated with warmer gas and is part of the H\,{\sc ii} region. 
Based on the temperature distribution and ionized emission, the bipolar nebula is clearly distinguished from the IRDC. 
This implies two distinct environments within the elongated filamentary structure.\\
$\bullet$ The position-velocity analysis of the gas kinematics of CO(3--2) suggests 
the signature of an expanding H\,{\sc ii} region with an expansion velocity of the gas to be $\sim$4.5 km s$^{-1}$;\\  
$\bullet$ In the position-velocity plot of CO(3--2), a noticeable velocity gradient has also been found toward the 
IRAS 17599$-$2148 position, indicating the presence of a molecular outflow. 
The exciting source of the outflow could be the driving source of the H\,{\sc ii} region and/or the IRc of the 6.7 GHz MME, 
however, due to coarse beam of CO data, we cannot conclude the powering source of the molecular outflow in this work.\\ 
$\bullet$ The analysis of the MIPSGAL, GLIMPSE, and UKIDSS-GPS photometry reveals 
a total of 474 YSOs. The YSOs clusters are spatially seen toward the condensations in the filamentary structure, 
revealing ongoing star formation. YSOs are also identified at the edges of the bipolar nebula. \\ 
$\bullet$ A point-like source is seen toward each compact radio continuum peak in the VLT/NACO images. 
There are no nebular features seen toward these sources in the NACO images. 
High-resolution spectroscopic study will be very helpful to confirm the ionizing sources of the radio peaks.\\\\ 
Based on our observed results, we conclude that the fragmentation has been occurred within the elongated filamentary structure, 
which produced several clumps (or a chain of dense clumps/cores) along the structure. Star formation activities are going on within these clumps.
These YSOs might have spontaneously formed due to processes not related to the expanding H\,{\sc ii} region. 
Massive stars were formed in one of the highest column density and massive clumps, and subsequently, an H\,{\sc ii} region was originated. 
This H\,{\sc ii} region has also been expanded in the surroundings and its ionizing feedback formed a bipolar nebula. 
At the edges of the bipolar nebula, four additional clumps are also associated with the YSOs clusters. 
The expanding H\,{\sc ii} region may also have triggered the star formation in the periphery of the nebula. 
\acknowledgments
We thank the anonymous referee for providing the constructive comments. 
The research work at Physical Research Laboratory is funded by the Department of Space, Government of India. 
This work is based on data obtained as part of the UKIRT Infrared Deep Sky Survey. This publication 
made use of data products from the Two Micron All Sky Survey (a joint project of the University of Massachusetts and 
the Infrared Processing and Analysis Center / California Institute of Technology, funded by NASA and NSF), archival 
data obtained with the {\it Spitzer} Space Telescope (operated by the Jet Propulsion Laboratory, California Institute 
of Technology under a contract with NASA). 
We thank the staff of the GMRT that made the radio observations possible.
The GMRT is run by the National Centre for  Radio Astrophysics of the Tata Institute of
Fundamental Research. 
IZ is supported by the Russian Foundation for Basic Research (RFBR). 
AL acknowledges the CONACYT(M\'{e}xico) grant CB-2012-01-1828-41. 
%\clearpage
%
%
\begin{deluxetable}{cccccccccccc}
\tablewidth{0pt} 
\tabletypesize{\scriptsize} 
\tablecaption{Summary of the properties of the identified {\it Herschel} clumps within the region probed in this paper 
using the {\it Herschel} data (see Figures~\ref{fig4}b and~\ref{fig4}c). Column 1 lists the IDs given to the clump. Table also contains 
positions, deconvolved effective radius (R$_{c}$), clump mass (M$_{clump}$), 
self-gravitating pressure ($P_{clump}$ $\approx$ $\pi G (M_{clump}/ \pi R_{c}^2$)$^2$), 
spherical free-fall time (t$_{ff,sph}$ $\approx$ $16.6 (R_{c}/pc)^{3/2} (M_{clump}/M_\odot)^{-1/2}$), peak velocity (V$_{peak}$), line-width ($\Delta V$), 
and virial mass (M$_{vir}$\,=126\,R$_{c}$\,$\Delta V{^2}$). \label{tab1}} 
\tablehead{ \colhead{ID} & \colhead{RA} & \colhead{Dec} & \colhead{R$_{c}$\tablenotemark{a}}& \colhead{M$_{clump}$}&\colhead{$P_{clump}$}& \colhead{t$_{ff,sph}$}& \colhead{V$_{peak}$}& \colhead{$\Delta V$}
& \colhead{M$_{vir}$}\\
\colhead{} &  \colhead{[J2000]} & \colhead{[J2000]} & \colhead{(pc)} &\colhead{($M_\odot$)} 
& \colhead{(10$^{-10}$ dynes\, cm$^{-2}$)}&\colhead{(Myr)}&\colhead{(km s$^{-1}$)}&\colhead{(km s$^{-1}$)}&\colhead{($M_\odot$)}}
\startdata 
    1	 &       18:03:04.7  & $-$21:48:13.0  & 2.0	 &    7023.7 & 29.2 & 0.39 &18.6\tablenotemark{b}&3.36\tablenotemark{b} &2830 \\
    2	  &      18:02:57.7  & $-$21:47:58.6  & 1.8	 &    4147.6 & 14.8 &0.47 &18.58\tablenotemark{b}&3.39\tablenotemark{b}&2625  \\
    3	 &       18:03:18.8  & $-$21:45:53.6  & 2.5	 &    5454.5 &6.9 &0.57&18.7\tablenotemark{c}&3.1\tablenotemark{c} &3047  \\
    4	  &      18:03:33.9  & $-$21:42:52.2  & 2.5	 &   6418.6 & 9.6  &0.52  &18.9\tablenotemark{c}&2.7\tablenotemark{c}&2306  \\
    5	  &      18:03:27.8  & $-$21:46:36.0  & 2.9	 &   6422    & 5.5& 0.60 &   --  & --  & --   \\
    6	 &       18:03:02.7 &  $-$21:45:38.9  & 1.4	 &   1192.6  &3.6& 0.66&--&-- &--   \\
    7	 &       18:02:56.6 &  $-$21:50:46.5  & 1.2	 &     763.8   &3.0 &  0.70&--&--&--   \\
    8	  &      18:03:00.6  & $-$21:50:18.7  & 1.4	 &    1046.2 &2.8& 0.70 &--&-- &--  \\
    9	 &       18:03:22.7  & $-$21:51:29.7  & 2.4	 &    3377.9 & 3.3& 0.70& --&--&--   \\
    10  &        18:02:40.6 &  $-$21:48:25.7 &  1.2	 &    776.8   & 2.3 &0.74& --&--&--  \\
    11   &       18:03:15.7 &  $-$21:50:47.4  & 1.8	 & 1678.5  &2.5&0.73& --&-- &--  \\
    12  &        18:02:56.7  & $-$21:46:06.6  & 0.9	 &   410    &2.1&0.76& --&-- &--  \\ 
\enddata  
\tablenotetext{a}{It is estimated using {\it clumpfind} algorithm.}
\tablenotetext{b}{NH$_{3}$(1,1) derived line parameters from \citet{wienen12}.}
\tablenotetext{c}{HCO$^{+}$(3--2) derived line parameters from \citet{shirley13}.}
\end{deluxetable}

\end{document}